\input harvmac.tex
\input epsf.tex

\def\figin{\epsfcheck\figin}\def\figins{\epsfcheck\figins}
\def\epsfcheck{\ifx\epsfbox\UnDeFiNeD
\message{(NO epsf.tex, FIGURES WILL BE IGNORED)}
\gdef\figin##1{\vskip2in}\gdef\figins##1{\hskip.5in}
\else\message{(FIGURES WILL BE INCLUDED)}%
\gdef\figin##1{##1}\gdef\figins##1{##1}\fi}
\def\DefWarn#1{}
\def\figinsert{\goodbreak\midinsert}
\def\ifig#1#2#3{\DefWarn#1\xdef#1{fig.~\the\figno}
\writedef{#1\leftbracket fig.\noexpand~\the\figno}%
\figinsert\figin{\centerline{#3}}\medskip\centerline{\vbox{\baselineskip12pt
\advance\hsize by -1truein\noindent\footnotefont{\bf
Fig.~\the\figno:} #2}}
\bigskip\endinsert\global\advance\figno by1}
\def \la {\langle}
\def \ra {\rangle}
\def \beq  {\begin{eqnarray}}
\def \eeq  {\end{eqnarray}}
\def \pa {\partial}
\def \eps {\epsilon}


\def\frac#1#2{{#1 \over #2}}
\def\text#1{#1}

%
\newbox\hdbox%
\newcount\hdrows%
\newcount\multispancount%
\newcount\ncase%
\newcount\ncols
\newcount\nrows%
\newcount\nspan%
\newcount\ntemp%
\newdimen\hdsize%
\newdimen\newhdsize%
\newdimen\parasize%
\newdimen\spreadwidth%
\newdimen\thicksize%
\newdimen\thinsize%
\newdimen\tablewidth%
\newif\ifcentertables%
\newif\ifendsize%
\newif\iffirstrow%
\newif\iftableinfo%
\newtoks\dbt%
\newtoks\hdtks%
\newtoks\savetks%
\newtoks\tableLETtokens%
\newtoks\tabletokens%
\newtoks\widthspec%
%
%
%
%
\tableinfotrue%
\catcode`\@=11
%
%
\def\tstrut{\vrule height3.1ex depth1.2ex width0pt}%
\def\and{\char`\&}
\def\tablerule{\noalign{\hrule height\thinsize depth0pt}}%
\thicksize=1.5pt
\thinsize=0.6pt
\def\thickrule{\noalign{\hrule height\thicksize depth0pt}}%
\def\ctr#1{\hfil\ #1\hfil}%
%
%
%
%
\tablewidth=-\maxdimen%
\spreadwidth=-\maxdimen%
\def\tabskipglue{0pt plus 1fil minus 1fil}%
%
%
\centertablestrue%
%
%
%
%
\parasize=4in%
\gdef\ARGS{########}
\gdef\headerARGS{####}
\def\@mpersand{&}
{\catcode`\|=13
\gdef\letbarzero{\let|0}
\gdef\letbartab{\def|{&&}}%
\gdef\letvbbar{\let\vb|}%
}
{\catcode`\&=4
\def\ampskip{&\omit\hfil&}
\catcode`\&=13
\let&0
\xdef\letampskip{\def&{\ampskip}}%
\gdef\letnovbamp{\let\novb&\let\tab&}
}
\def\begintable{
   \begingroup%
   \catcode`\|=13\letbartab\letvbbar%
   \catcode`\&=13\letampskip\letnovbamp%
   \def\multispan##1{
      \omit \mscount##1%
      \multiply\mscount\tw@\advance\mscount\m@ne%
      \loop\ifnum\mscount>\@ne \sp@n\repeat%
   }
   \def\|{%
      &\omit\widevline&%
   }%
   \ruledtable
}
\long\def\ruledtable#1\endtable{%
%
%
%
   \offinterlineskip
   \tabskip 0pt
   \def\widevline{\vrule width\thicksize}
   \def\endrow{\@mpersand\omit\hfil\crnorm\@mpersand}%
   \def\crthick{\@mpersand\crnorm\thickrule\@mpersand}%
   \def\crthickneg##1{\@mpersand\crnorm\thickrule
          \noalign{{\skip0=##1\vskip-\skip0}}\@mpersand}%
   \def\crnorule{\@mpersand\crnorm\@mpersand}%
   \def\crnoruleneg##1{\@mpersand\crnorm
          \noalign{{\skip0=##1\vskip-\skip0}}\@mpersand}%
   \let\nr=\crnorule
   \def\endtable{\@mpersand\crnorm\thickrule}%
   \let\crnorm=\cr
%
%
   \edef\cr{\@mpersand\crnorm\tablerule\@mpersand}%
   \def\crneg##1{\@mpersand\crnorm\tablerule
          \noalign{{\skip0=##1\vskip-\skip0}}\@mpersand}%
   \let\ctneg=\crthickneg
   \let\nrneg=\crnoruleneg
   \the\tableLETtokens
%
%
   \tabletokens={&#1}
%
%
   \countROWS\tabletokens\into\nrows%
   \countCOLS\tabletokens\into\ncols%
%
%
   \advance\ncols by -1%
   \divide\ncols by 2%
   \advance\nrows by 1%
%
%
   \iftableinfo %
      \immediate\write16{[Nrows=\the\nrows, Ncols=\the\ncols]}%
   \fi%
%
%
   \ifcentertables
      \ifhmode \par\fi
      \line{
      \hss
   \else %
      \hbox{%
   \fi
      \vbox{%
         \makePREAMBLE{\the\ncols}
         \edef\next{\preamble}
         \let\preamble=\next
         \makeTABLE{\preamble}{\tabletokens}
      }
      \ifcentertables \hss}\else }\fi
   \endgroup
   \tablewidth=-\maxdimen
   \spreadwidth=-\maxdimen
}
\def\makeTABLE#1#2{
   {
   \let\ifmath0
   \let\header0
   \let\multispan0
%
%
   \ncase=0%
   \ifdim\tablewidth>-\maxdimen \ncase=1\fi%
   \ifdim\spreadwidth>-\maxdimen \ncase=2\fi%
   \relax
%
   \ifcase\ncase %
      \widthspec={}%
   \or %
      \widthspec=\expandafter{\expandafter t\expandafter o%
                 \the\tablewidth}%
   \else %
      \widthspec=\expandafter{\expandafter s\expandafter p\expandafter r%
                 \expandafter e\expandafter a\expandafter d%
                 \the\spreadwidth}%
   \fi %
   \xdef\next{
      \halign\the\widthspec{%
      #1
      \noalign{\hrule height\thicksize depth0pt}
      \the#2\endtable
%
      }
   }
   }
   \next
}
\def\makePREAMBLE#1{
   \ncols=#1
   \begingroup
   \let\ARGS=0
   \edef\xtp{\widevline\ARGS\tabskip\tabskipglue%
   &\ctr{\ARGS}\tstrut}
   \advance\ncols by -1
   \loop
      \ifnum\ncols>0 %
      \advance\ncols by -1%
      \edef\xtp{\xtp&\vrule width\thinsize\ARGS&\ctr{\ARGS}}%
   \repeat
   \xdef\preamble{\xtp&\widevline\ARGS\tabskip0pt%
   \crnorm}
   \endgroup
}
\def\countROWS#1\into#2{
   \let\countREGISTER=#2%
   \countREGISTER=0%
   \expandafter\ROWcount\the#1\endcount%
}%
\def\ROWcount{%
   \afterassignment\subROWcount\let\next= %
}%
\def\subROWcount{%
   \ifx\next\endcount %
      \let\next=\relax%
   \else%
      \ncase=0%
      \ifx\next\cr %
         \global\advance\countREGISTER by 1%
         \ncase=0%
      \fi%
      \ifx\next\endrow %
         \global\advance\countREGISTER by 1%
         \ncase=0%
      \fi%
      \ifx\next\crthick %
         \global\advance\countREGISTER by 1%
         \ncase=0%
      \fi%
      \ifx\next\crnorule %
         \global\advance\countREGISTER by 1%
         \ncase=0%
      \fi%
      \ifx\next\crthickneg %
         \global\advance\countREGISTER by 1%
         \ncase=0%
      \fi%
      \ifx\next\crnoruleneg %
         \global\advance\countREGISTER by 1%
         \ncase=0%
      \fi%
      \ifx\next\crneg %
         \global\advance\countREGISTER by 1%
         \ncase=0%
      \fi%
      \ifx\next\header %
         \ncase=1%
      \fi%
      \relax%
      \ifcase\ncase %
         \let\next\ROWcount%
      \or %
         \let\next\argROWskip%
      \else %
      \fi%
   \fi%
   \next%
}
\def\counthdROWS#1\into#2{%
\dvr{10}%
   \let\countREGISTER=#2%
   \countREGISTER=0%
\dvr{11}%
\dvr{13}%
   \expandafter\hdROWcount\the#1\endcount%
\dvr{12}%
}%
\def\hdROWcount{%
   \afterassignment\subhdROWcount\let\next= %
}%
\def\subhdROWcount{%
   \ifx\next\endcount %
      \let\next=\relax%
   \else%
      \ncase=0%
      \ifx\next\cr %
         \global\advance\countREGISTER by 1%
         \ncase=0%
      \fi%
      \ifx\next\endrow %
         \global\advance\countREGISTER by 1%
         \ncase=0%
      \fi%
      \ifx\next\crthick %
         \global\advance\countREGISTER by 1%
         \ncase=0%
      \fi%
      \ifx\next\crnorule %
         \global\advance\countREGISTER by 1%
         \ncase=0%
      \fi%
      \ifx\next\header %
         \ncase=1%
      \fi%
\relax%
      \ifcase\ncase %
         \let\next\hdROWcount%
      \or%
         \let\next\arghdROWskip%
      \else %
      \fi%
   \fi%
   \next%
}%
{\catcode`\|=13\letbartab
\gdef\countCOLS#1\into#2{%
   \let\countREGISTER=#2%
   \global\countREGISTER=0%
   \global\multispancount=0%
   \global\firstrowtrue
   \expandafter\COLcount\the#1\endcount%
   \global\advance\countREGISTER by 3%
   \global\advance\countREGISTER by -\multispancount
}%
\gdef\COLcount{%
   \afterassignment\subCOLcount\let\next= %
}%
{\catcode`\&=13%
\gdef\subCOLcount{%
   \ifx\next\endcount %
      \let\next=\relax%
   \else%
      \ncase=0%
      \iffirstrow
         \ifx\next& %
            \global\advance\countREGISTER by 2%
            \ncase=0%
         \fi%
         \ifx\next\span %
            \global\advance\countREGISTER by 1%
            \ncase=0%
         \fi%
         \ifx\next| %
            \global\advance\countREGISTER by 2%
            \ncase=0%
         \fi
         \ifx\next\|
            \global\advance\countREGISTER by 2%
            \ncase=0%
         \fi
         \ifx\next\multispan
            \ncase=1%
            \global\advance\multispancount by 1%
         \fi
         \ifx\next\header
            \ncase=2%
         \fi
         \ifx\next\cr       \global\firstrowfalse \fi
         \ifx\next\endrow   \global\firstrowfalse \fi
         \ifx\next\crthick  \global\firstrowfalse \fi
         \ifx\next\crnorule \global\firstrowfalse \fi
         \ifx\next\crnoruleneg \global\firstrowfalse \fi
         \ifx\next\crthickneg  \global\firstrowfalse \fi
         \ifx\next\crneg       \global\firstrowfalse \fi
      \fi
\relax
      \ifcase\ncase %
         \let\next\COLcount%
      \or %
         \let\next\spancount%
      \or %
         \let\next\argCOLskip%
      \else %
      \fi %
   \fi%
   \next%
}%
\gdef\argROWskip#1{%
   \let\next\ROWcount \next%
}
\gdef\arghdROWskip#1{%
   \let\next\ROWcount \next%
}
\gdef\argCOLskip#1{%
   \let\next\COLcount \next%
}
}
}
\def\spancount#1{
   \nspan=#1\multiply\nspan by 2\advance\nspan by -1%
   \global\advance \countREGISTER by \nspan
   \let\next\COLcount \next}%
\def\dvr#1{\relax}%
\def\header#1{%
\dvr{1}{\let\cr=\@mpersand%
\hdtks={#1}%
\counthdROWS\hdtks\into\hdrows%
\advance\hdrows by 1%
\ifnum\hdrows=0 \hdrows=1 \fi%
\dvr{5}\makehdPREAMBLE{\the\hdrows}%
\dvr{6}\getHDdimen{#1}%
{\parindent=0pt\hsize=\hdsize{\let\ifmath0%
\xdef\next{\valign{\headerpreamble #1\crnorm}}}\dvr{7}\next\dvr{8}%
}%
}\dvr{2}}
\def\makehdPREAMBLE#1{
\dvr{3}%
\hdrows=#1
{
\let\headerARGS=0%
\let\cr=\crnorm%
\edef\xtp{\vfil\hfil\hbox{\headerARGS}\hfil\vfil}%
\advance\hdrows by -1
\loop
\ifnum\hdrows>0%
\advance\hdrows by -1%
\edef\xtp{\xtp&\vfil\hfil\hbox{\headerARGS}\hfil\vfil}%
\repeat%
\xdef\headerpreamble{\xtp\crcr}%
}
\dvr{4}}
\def\getHDdimen#1{%
\hdsize=0pt%
\getsize#1\cr\end\cr%
}
\def\getsize#1\cr{%
\endsizefalse\savetks={#1}%
\expandafter\lookend\the\savetks\cr%
\relax \ifendsize \let\next\relax \else%
\setbox\hdbox=\hbox{#1}\newhdsize=1.0\wd\hdbox%
\ifdim\newhdsize>\hdsize \hdsize=\newhdsize \fi%
\let\next\getsize \fi%
\next%
}%
\def\lookend{\afterassignment\sublookend\let\looknext= }%
\def\sublookend{\relax%
\ifx\looknext\cr %
\let\looknext\relax \else %
   \relax
   \ifx\looknext\end \global\endsizetrue \fi%
   \let\looknext=\lookend%
    \fi \looknext%
}%
%
%
\def\tablelet#1{%
   \tableLETtokens=\expandafter{\the\tableLETtokens #1}%
}%
\catcode`\@=12
%


\lref\KorchemskySpin{ A.~V.~Belitsky, A.~S.~Gorsky and G.~P.~Korchemsky,
  Nucl.\ Phys.\  B {\bf 748}, 24 (2006)
  [arXiv:hep-th/0601112].
}

\lref\HitchinVP{
  N.~J.~Hitchin,
  Proc.\ Lond.\ Math.\ Soc.\  {\bf 55}, 59 (1987).
}

\lref\ZamolodchikovET{
  A.~B.~Zamolodchikov,
  Phys.\ Lett.\  B {\bf 253}, 391 (1991).
}

\lref\FendleyVE{
  P.~Fendley and K.~A.~Intriligator,
  Nucl.\ Phys.\  B {\bf 372}, 533 (1992)
  [arXiv:hep-th/9111014].
}

\lref\FendleyXN{
  P.~Fendley,
  Nucl.\ Phys.\  B {\bf 374}, 667 (1992)
  [arXiv:hep-th/9109021].
}

\lref\ZamolodchikovUW{
  A.~B.~Zamolodchikov,
  Nucl.\ Phys.\  B {\bf 432}, 427 (1994)
  [arXiv:hep-th/9409108].
}

\lref\AGM{
  L.~F.~Alday, D.~Gaiotto and J.~Maldacena,
  arXiv:0911.4708 [hep-th].
}
\lref\octagon{
  L.~F.~Alday and J.~Maldacena,
  JHEP {\bf 0911}, 082 (2009)
  [arXiv:0904.0663 [hep-th]].
}

\lref\AMSV{
  L.~F.~Alday, J.~Maldacena, A.~Sever and P.~Vieira,
  arXiv:1002.2459 [hep-th].
}

\lref\GKP{
  S.~S.~Gubser, I.~R.~Klebanov and A.~M.~Polyakov,
  Nucl.\ Phys.\  B {\bf 636}, 99 (2002)
  [arXiv:hep-th/0204051].
}

\lref\AMtwo{ L.~F.~Alday and J.~M.~Maldacena,
  JHEP {\bf 0711}, 019 (2007)
  [arXiv:0708.0672 [hep-th]].
}

\lref\Basso{B. Basso, to appear.}

\lref\AlexandrovPP{
  S.~Alexandrov and P.~Roche,
  arXiv:1003.3964 [hep-th].
}
\lref\NekrasovRC{
  N.~A.~Nekrasov and S.~L.~Shatashvili,
  arXiv:0908.4052 [hep-th].
}

\lref\Kruczenski{
  M.~Kruczenski,
  JHEP {\bf 0212}, 024 (2002)
  [arXiv:hep-th/0210115].
}
\lref\KorchemskyMar{
  G.~P.~Korchemsky and G.~Marchesini,
  Nucl.\ Phys.\  B {\bf 406}, 225 (1993)
  [arXiv:hep-ph/9210281].
}

\lref\GMNtwo{
  D.~Gaiotto, G.~W.~Moore and A.~Neitzke,
  arXiv:0907.3987 [hep-th].
}
\lref\BMN{
  D.~E.~Berenstein, J.~M.~Maldacena and H.~S.~Nastase,
  JHEP {\bf 0204}, 013 (2002)
  [arXiv:hep-th/0202021].
}

\lref\GMNone{
  D.~Gaiotto, G.~W.~Moore and A.~Neitzke,
  arXiv:0807.4723 [hep-th].
}

\lref\SokWard{ J.~M.~Drummond, J.~Henn, G.~P.~Korchemsky and E.~Sokatchev,
  Nucl.\ Phys.\  B {\bf 826}, 337 (2010)
  [arXiv:0712.1223 [hep-th]].
}
\lref\BDS{
  Z.~Bern, L.~J.~Dixon and V.~A.~Smirnov,
  Phys.\ Rev.\  D {\bf 72}, 085001 (2005)
  [arXiv:hep-th/0505205].
}

\lref\ConformalQCD{
  V.~M.~Braun, G.~P.~Korchemsky and D.~Mueller,
  Prog.\ Part.\ Nucl.\ Phys.\  {\bf 51}, 311 (2003)
  [arXiv:hep-ph/0306057].
}

\lref\brandhuber{
  A.~Brandhuber, P.~Heslop and G.~Travaglini,
  Nucl.\ Phys.\  B {\bf 794}, 231 (2008)
  [arXiv:0707.1153 [hep-th]].
}
\lref\BES{N.~Beisert, B.~Eden and M.~Staudacher,
  J.\ Stat.\ Mech.\  {\bf 0701}, P021 (2007)
  [arXiv:hep-th/0610251].
}

\lref\Zarembo{
 K.~Zarembo,
  JHEP {\bf 0904}, 135 (2009)
  [arXiv:0903.1747 [hep-th]].
}

\lref\BPZ{
  A.~A.~Belavin, A.~M.~Polyakov and A.~B.~Zamolodchikov,
  Nucl.\ Phys.\  B {\bf 241}, 333 (1984).
}

\lref\FL{
  N.~Beisert and M.~Staudacher,
  Nucl.\ Phys.\  B {\bf 670} (2003) 439
  [arXiv:hep-th/0307042].
  $\bullet$ 
  N.~Beisert,
  Phys.\ Rept.\  {\bf 405}, 1 (2005)
  [arXiv:hep-th/0407277].
  }

  \lref\ZhangTR{
  J.~H.~Zhang,
  arXiv:1004.1606 [hep-th].
}

\lref\AldayHR{
  L.~F.~Alday and J.~M.~Maldacena,
  JHEP {\bf 0706}, 064 (2007)
  [arXiv:0705.0303 [hep-th]].
}

\lref\DrummondAUA{
  J.~M.~Drummond, G.~P.~Korchemsky and E.~Sokatchev,
  Nucl.\ Phys.\  B {\bf 795}, 385 (2008)
  [arXiv:0707.0243 [hep-th]].
}

\lref\BrandhuberYX{
  A.~Brandhuber, P.~Heslop and G.~Travaglini,
  Nucl.\ Phys.\  B {\bf 794}, 231 (2008)
  [arXiv:0707.1153 [hep-th]].
}

\lref\DrummondAQ{
  J.~M.~Drummond, J.~Henn, G.~P.~Korchemsky and E.~Sokatchev,
  Nucl.\ Phys.\  B {\bf 815}, 142 (2009)
  [arXiv:0803.1466 [hep-th]].
}

\lref\AldayHE{
  L.~F.~Alday and J.~Maldacena,
  JHEP {\bf 0711}, 068 (2007)
  [arXiv:0710.1060 [hep-th]].
}

\lref\DrummondBM{
  J.~M.~Drummond, J.~Henn, G.~P.~Korchemsky and E.~Sokatchev,
  Phys.\ Lett.\  B {\bf 662}, 456 (2008)
  [arXiv:0712.4138 [hep-th]].
}

\lref\BernAP{
  Z.~Bern, L.~J.~Dixon, D.~A.~Kosower, R.~Roiban, M.~Spradlin, C.~Vergu and A.~Volovich,
  Phys.\ Rev.\  D {\bf 78}, 045007 (2008)
  [arXiv:0803.1465 [hep-th]].
}

\lref\AnastasiouKNA{
  C.~Anastasiou, A.~Brandhuber, P.~Heslop, V.~V.~Khoze, B.~Spence and G.~Travaglini,
  JHEP {\bf 0905}, 115 (2009)
  [arXiv:0902.2245 [hep-th]].
}

\lref\DelDucaZG{
  V.~Del Duca, C.~Duhr and V.~A.~Smirnov,
  JHEP {\bf 1005}, 084 (2010)
  [arXiv:1003.1702 [hep-th]].
}

\lref\HodgesHK{
  A.~Hodges,
  arXiv:0905.1473 [hep-th].
}

\lref\McCoyCD{
  B.~M.~McCoy, C.~A.~Tracy and T.~T.~Wu,
  J.\ Math.\ Phys.\  {\bf 18}, 1058 (1977).
}

\lref\GaiottoHG{
  D.~Gaiotto, G.~W.~Moore and A.~Neitzke,
  arXiv:0907.3987 [hep-th].
}
\lref\AldayZY{
  L.~F.~Alday, B.~Eden, G.~P.~Korchemsky, J.~Maldacena and E.~Sokatchev,
  arXiv:1007.3243 [hep-th].
}

\lref\AldayKU{
  L.~F.~Alday, D.~Gaiotto, J.~Maldacena, A.~Sever and P.~Vieira,
  arXiv:1006.2788 [hep-th].
}

\lref\AldayVH{
  L.~F.~Alday, J.~Maldacena, A.~Sever and P.~Vieira,
  arXiv:1002.2459 [hep-th].
}

\lref\AldayDV{
  L.~F.~Alday, D.~Gaiotto and J.~Maldacena,
  arXiv:0911.4708 [hep-th].
}

\lref\AldayYN{
  L.~F.~Alday and J.~Maldacena,
  JHEP {\bf 0911}, 082 (2009)
  [arXiv:0904.0663 [hep-th]].
}

\lref\AldayGA{
  L.~F.~Alday and J.~Maldacena,
  arXiv:0903.4707 [hep-th].
}

\lref\AldayZZA{
  L.~F.~Alday and J.~Maldacena,
  AIP Conf.\ Proc.\  {\bf 1031}, 43 (2008).
}

\lref\AldayHE{
  L.~F.~Alday and J.~Maldacena,
  JHEP {\bf 0711}, 068 (2007)
  [arXiv:0710.1060 [hep-th]].
}

\lref\AldayMF{
  L.~F.~Alday and J.~M.~Maldacena,
  JHEP {\bf 0711}, 019 (2007)
  [arXiv:0708.0672 [hep-th]].
}

\lref\AldayHR{
  L.~F.~Alday and J.~M.~Maldacena,
  JHEP {\bf 0706}, 064 (2007)
  [arXiv:0705.0303 [hep-th]].
}

\lref\LukyanovRN{
  S.~L.~Lukyanov and A.~B.~Zamolodchikov,
  JHEP {\bf 1007}, 008 (2010)
  [arXiv:1003.5333 [math-ph]].
}

\lref\SeibergEB{
  N.~Seiberg,
  Prog.\ Theor.\ Phys.\ Suppl.\  {\bf 102}, 319 (1990).
}

\lref\EdenZZ{
  B.~Eden, G.~P.~Korchemsky and E.~Sokatchev,
  arXiv:1007.3246 [hep-th].
}

\lref\AldayZY{
  L.~F.~Alday, B.~Eden, G.~P.~Korchemsky, J.~Maldacena and E.~Sokatchev,
  arXiv:1007.3243 [hep-th].
}

\lref\BanadosWN{
  M.~Banados, C.~Teitelboim and J.~Zanelli,
  Phys.\ Rev.\ Lett.\  {\bf 69}, 1849 (1992)
  [arXiv:hep-th/9204099].
}

\lref\YangRM{
  C.~N.~Yang and C.~P.~Yang,
  J.\ Math.\ Phys.\  {\bf 10}, 1115 (1969).
}

\lref\ZamolodchikovCF{
  A.~B.~Zamolodchikov,
  Nucl.\ Phys.\  B {\bf 342}, 695 (1990).
}

\lref\HofmanAR{
  D.~M.~Hofman and J.~Maldacena,
  JHEP {\bf 0805}, 012 (2008)
  [arXiv:0803.1467 [hep-th]].
}

\lref\DoreyZJ{
  N.~Dorey and B.~Vicedo,
  JHEP {\bf 0607}, 014 (2006)
  [arXiv:hep-th/0601194].
}

\lref\SteifZM{
  A.~R.~Steif,
  Phys.\ Rev.\  D {\bf 53}, 5521 (1996)
  [arXiv:hep-th/9504012].
}

\lref\KazakovQF{
  V.~A.~Kazakov, A.~Marshakov, J.~A.~Minahan and K.~Zarembo,
  JHEP {\bf 0405}, 024 (2004)
  [arXiv:hep-th/0402207].
}

\lref\PolchinskiJW{
  J.~Polchinski and M.~J.~Strassler,
  JHEP {\bf 0305}, 012 (2003)
  [arXiv:hep-th/0209211].
}
\lref\YangAS{
  G.~Yang,
  arXiv:1004.3983 [hep-th].
}

\lref\DeVegaXC{
  H.~J.~De Vega and N.~G.~Sanchez,
  Phys.\ Rev.\  D {\bf 47}, 3394 (1993).
}

\lref\PohlmeyerNB{
  K.~Pohlmeyer,
  Commun.\ Math.\ Phys.\  {\bf 46}, 207 (1976).
}

\lref\vanNeervenJA{
  W.~L.~van Neerven,
  Z.\ Phys.\  C {\bf 30}, 595 (1986).
}

\lref\Kirillov{
A.N.Kirillov, Journal of Mathematical Sciences, Volume 47, Number 2 / October, 1989}

\Title{\vbox{\baselineskip12pt }}
{\vbox{\centerline{Form factors at strong coupling}
\vskip .05in
\centerline{via a Y-system}
}}
\bigskip
\centerline{  Juan Maldacena$^a$ and Alexander Zhiboedov$^b$ }
\bigskip

\centerline{ \it  $^a$School of Natural Sciences, Institute for Advanced Study} \centerline{\it Princeton, NJ 08540, USA}

\centerline{\it $^b$ Jadwin Hall, Princeton University
 }
\centerline{ \it Princeton, NJ 08540, USA}

\vskip .3in \noindent
We compute form factors in   planar ${\cal N}=4$ Super Yang-Mills at strong coupling.
Namely we consider the overlap between an operator insertion and $2n$ gluons.
Through the gauge/string duality these are given by minimal surfaces in $AdS$ space. The surfaces end on an
 infinite periodic sequence
of null segments at the boundary of $AdS$. We consider   surfaces that can be embedded in $AdS_{3}$.
We derive   set of functional equations for the cross
ratios as functions of the spectral parameter. These equations are of the form of a  Y-system.
The integral form of the Y-system has Thermodynamics
Bethe Ansatz form. The area is given by the free energy of the TBA system or critical value of Yang-Yang functional.
We consider a restricted set of operators which
have small conformal  dimension
compared to $\sqrt{\lambda}$.


 \Date{ }


\listtoc\writetoc
\vskip .5in \noindent

\newsec{Introduction}
\noindent

Recently there has been considerable progress in the calculation of
 light-like Wilson loops both at weak and strong coupling. These are Lorentzian objects
that depend on a finite number of parameters, namely positions of the cusps.
One of the main motivations to calculate such objects was the fact that they describe color-ordered
scattering amplitudes in ${\cal N}=4$ SYM \refs{\AldayHR, \DrummondAUA, \BrandhuberYX, \BernAP, \AnastasiouKNA, \DelDucaZG}. The momenta of the particles, $k_{i}^{\mu}$, define the contour by the simple rule
$x_{i}^{\mu} - x_{i+1}^{\mu} = k^{\mu}_{i}$. However, very soon it became clear that light-like Wilson loops are related to broader
class of objects, namely to form factors, non-supersymmetric Wilson loops and correlation functions \refs{\AldayHE,\AldayKU,\AldayZY,\EdenZZ}.

At strong coupling,  the problem can be solved by  using   the integrability of classical strings on
$AdS$  \refs{\AldayYN, \AldayDV, \AldayVH}.
The key steps were: first, introduce a  family of flat connections parameterized by an arbitrary spectral
parameter ${\cal A}(\zeta=e^{\theta})$. With these flat connections one can consider the
 flat section problem which allows one to introduce a set
of the spectral parameter dependent cross ratios $Y_{k}(\theta)$. Then one finds a set of functional equations which
constraints the $\theta$ dependence of Y-functions. The system is specified with boundary conditions at $\theta \to \pm \infty$ which
are obtained through a WKB analysis of the flat section problem.
Then one can derive
 an integral form of these equations.
  These are of a   TBA form \refs{\YangRM, \ZamolodchikovCF, \GMNone, \GaiottoHG}. The non-trivial part of the area is given by the free energy of that TBA system which depends on a
   set of mass parameters in terms of which the kinematics is encoded.
    It is  also
     convenient to have an
      expression for the area that depends only on the  kinematic information,
       namely cross ratios. This can be obtained by
       solving for the
        mass parameters in terms of the physical cross ratios. The
        final area is given   by the critical value of Yang-Yang functional \refs{\AldayKU}.

In this paper we consider a more general problem which corresponds to the
additional insertion of a closed string state on the worldsheet.
At the insertion point, $z_{0}$, the connection ${\cal A}(z_{0})$ exhibits specific singularity.
The inserted state is specified by the monodromy $\Omega(\zeta)$ of the flat connection around the insertion point
 \refs{\KazakovQF,\DoreyZJ}.

Physically, we are studying   processes involving a local operator and a fixed number of particles.
Such objects are called form factors.
The operator creates a state which is then projected on to a state with a definite number of gluons.
 This is a problem that is common in QFT.
   For example, one can have in mind all processes like $e^{+}e^{-} \to {\rm jets}$ to lowest order in $\alpha_{em}$ but to
all orders in $\alpha_{s}$. Here the electromagnetic current creates a QCD state which we want to study.

\ifig\Periodic{In (a) we show a  worldsheet with   topology of a disc. It has
 open string insertions at the boundary corresponding to the gluons
 and a closed string insertion corresponding to an operator with momentum   $q^{\mu}$.
 In (b) we show the kinematics of the process.
Each gluon has a light-like momentum. The operator has an
arbitrary momentum, for convenience we choose it to be spacelike and collinear to ${\rm x}^{1}$ axis.
In (c) we show the periodic configuration which arises from the configuration (b) after we perform the T-duality transformation.
Note that the doted line in (b) is {\it not} part of the contour of the Wilson loop.  } {\epsfxsize3.5in\epsfbox{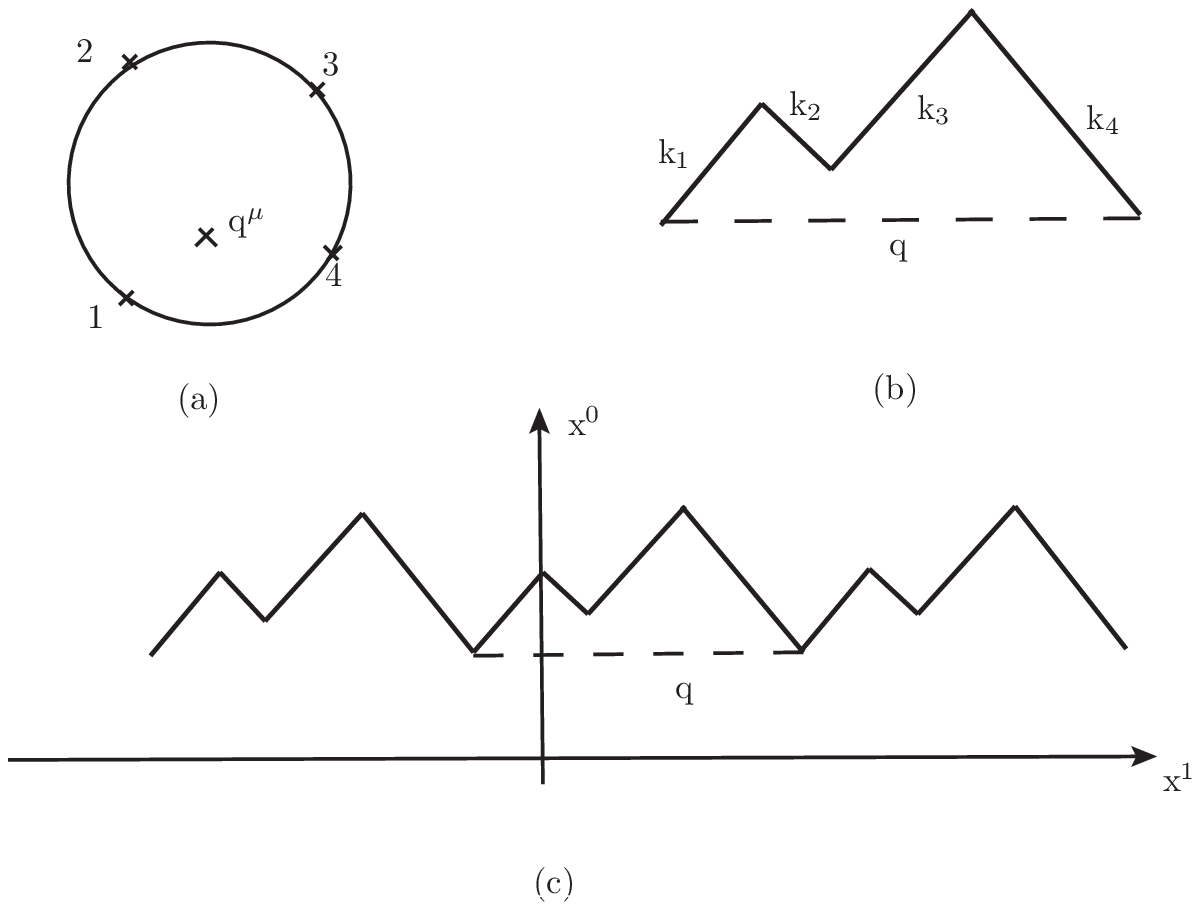}}

At strong coupling the problem involves computing the area of a certain minimal surface. The geometric
problem is obtained after performing a certain T-duality, as explained in \AldayHE .
 The surfaces end on an infinite periodic sequence of
null segments at the boundary of $AdS$ space. The period
is defined by the momentum, $q^\mu$,  of the inserted operator.  The shape of the polygon within each period
is defined by momenta of gluons, see \Periodic . Another manifestation of the presence of the operator
 comes from the fact that
we consider   solutions that go to the horizon.

We consider
surfaces that can be embedded in $AdS_{3}$. This amounts to considering
 $R^{1,1}$ kinematics in dual field theory language\foot{Remember that we limit the kinematics of the external gluons.
  Gluons propagating in the loops live in $R^{1,3}$.}.
We extend the approach of \refs{\AldayVH} to   cases with general monodromy
and derive a set of functional equations for the conformal cross as functions of the
spectral parameter. Then we study several cases when the monodromy is   spectral parameter independent.
In the case of trivial monodromy they reduce to the Y-system for scattering amplitudes.

We then consider form factors defined via
\eqn\ff{
\la \{k_{i}^{in}\}|{\cal O}(q)|\{k_{j}^{out}\} \ra = \int d^{4}x e^{i q x} \la \{k_{i}^{in}\}|{\cal O}(x)|\{k_{j}^{out}\} \ra.
}
We consider operators with conformal dimensions
small compared to $\sqrt{\lambda}$ \foot{Examples of such operators are the stress  tensor, the  R-currents, any BPS operator,
low level massive string states with dimensions $\propto \lambda^{1/4}$. This excludes operators corresponding
to semiclassical string states that have energies going like $\lambda^{1/2}$.}.
The problem boils down to studying strings in massless BTZ black hole geometry \refs{\BanadosWN}.
This geometry is simply $AdS_{3}$ with the identification $x \sim x + q$ where $x$ is space Poincare coordinate see \Periodic.
Originally the prescription for calculating such objects was found in \refs{\AldayHE}. At weak coupling these objects were considered in
 \refs{\vanNeervenJA}.

Here we analyze this case in detail and find all pieces of the area. The most complicated one is given by the
free energy of the TBA system   or as   the critical value of Yang-Yang functional.
We get a result that is independent of the operator that we insert (as long as they have small dimension compared to
$\sqrt{\lambda}$).
  One can understand it qualitatively as follows:
  at strong coupling there is copious production of very low energy gluons, so the emission of small number of gluons is
  equally
  strongly disfavored for all operators \refs{\PolchinskiJW,\HofmanAR}. The dependence on the operator, and on the
  polarization of the gluons, should reappear at one loop in the $1/\sqrt{\lambda}$ expansion.

Our paper is organized as follows. In section two we review \AldayHE .
In section three we derive  the Y-system for
a general monodromy. This is a general discussion which applies to cases that are more
general than what we discuss later.
In section four we analyze the case when the monodromy does not depend
on the spectral parameter. In section five
we concentrate on Wilson loops in a massless BTZ background and derive integral form
of the equations which are of the  TBA form. The most non-trivial part of the area is
given by the free energy of the corresponding TBA system or critical value of Yang-Yang functional.
In section six we analyze all pieces
of the area. In section seven we consider exact solutions of the Y-system which allows us to
check our formulas. In addition we derive
the explicit answer for the case of an operator going into four gluons.
We end with   conclusions. We include   several appendices with useful
technical details. In particular,  we explain how one can compute the area when the number
of gluons is proportional to $4$ by taking the double soft limit, both for amplitudes and for the case of form factor (this was
also considered in \YangAS ).

\newsec{Short review of the strong coupling prescription for computing form factors}

Here we briefly recall the prescription for calculation of the form factors at strong coupling \AldayHE.

We are working in  the $AdS_5$ space describing the gravity dual to ${\cal N}=4$ SYM with the metric
\eqn\Ads{
d s^2 = {d x^2 + d z^2 \over z^2}
}
where $x^{\mu}$ are the coordinates of the $R^{1,3}$ space where the field theory lives.

Since scattering amplitudes in CFT's are ill-defined one needs to introduce IR
regulator. A convenient   regulator consists of D-branes which are located at $z_{IR}$. Gluons are
open strings ending on them.
 Removing the regulator corresponds to sending $z_{IR} \to \infty$.
After we introduced the regulator we can consider complex
 classical solutions of the
 string equations of motion whose boundary conditions
 in the past and future infinity are set at $z= z_{IR}$ where asymptotic gluons live.
 In addition, the worldsheet reaches $z=0$ where the operator is inserted.

To describe these classical solutions it is very convenient to perform a  T-duality transformation along the
four worldvolume directions and to make the change of variables $r = {1 \over z}$.
This transformation leaves the metric invariant
and exchanges the momenta by ``windings''.

In the new coordinates the gluon states are represented by a sequence of light-like lines $k_{i}^{\mu}$ at $r=0$. However because of
the presence of the operator this sequence is not closed, namely we have $\sum_{i=1}^{2n} k_{i}^{\mu} = q^{\mu}$ where $q^{\mu}$
is the operator momentum. The operator is represented by a closed string so we should identify the ends of the
sequence which is the same as compactifying
 the coordinate along $q^{\mu}$  and considering a   closed string winding this coordinate.
Equivalently we can unfold this picture and consider an infinite periodic set of light-like segments given by momenta $k^{\mu}_{i}$.
For the case of $AdS_3$ which will be in the focus of the paper this procedure is illustrated in \Periodic.
 The operator which was inserted
 at $z=0$, now leads to a string worldsheet which goes to $r = \infty$.

After one finds the solution that minimizes the area with given boundary condition the form factor takes the form
\eqn\Formf{
\la \{k_{i}^{in}\}|{\cal O}(q)|\{k_{j}^{out}\} \ra =
e^{- \frac{\sqrt{\lambda}}{2 \pi} ({\rm Area})_{T}} {\rm F}_{1} (1 + {1 \over \sqrt{\lambda}} {\rm F}_{2} + \cdots)
}
where the leading term $({\rm Area})_{T}$ is the area of one period and its
 computation is the main topic of this paper. ${\rm F}_1$ is
the one loop correction and would contain information about the
polarizations and the particular operator we are considering. ${\rm F}_2$ is the
two loop correction in the $1/\sqrt{\lambda}$ expansion.

\newsec{Derivation of the Y-system for general monodromy}

We consider minimal surfaces that can be embedded in $AdS_{3}$ subspace of $AdS_{5}$.
From the field theory side it corresponds to   gluons and operators  with momenta lying in an  $R^{1,1}$ subspace
of $R^{1,3}$. Throughout this  paper we consider a  problem with $2n$ gluons.
We first consider $n$ odd,
which simplifies the consideration. We then  get the answer for $n$ even by taking double soft limit at
the very end.

\subsec{Preliminaries}

We are interested in the classical dynamics of strings in $AdS_{3}$ space.
$AdS_{3}$ can be conveniently written as the following surface in $R^{2,2}$
\eqn\AdSthree{
Z_{-1}^2 + Z_{0}^2 - Z^2_{1}-Z_{2}^{2} = 1.
}
We use the Poincare coordinates
\eqn\Poincare{
x^{\pm} = \frac{Z_{1} \pm Z_{0}}{Z_{-1} + Z_{2}}, \quad r = \frac{1}{Z_{-1} + Z_{2}}
}
to define the kinematics of the process.

\ifig\Label{The labeling of cusps at the boundary of $AdS_{3}$. The $q$ stands for the operator momentum and it is the period of the polygon at the same time.
 }{\epsfxsize3.0in\epsfbox{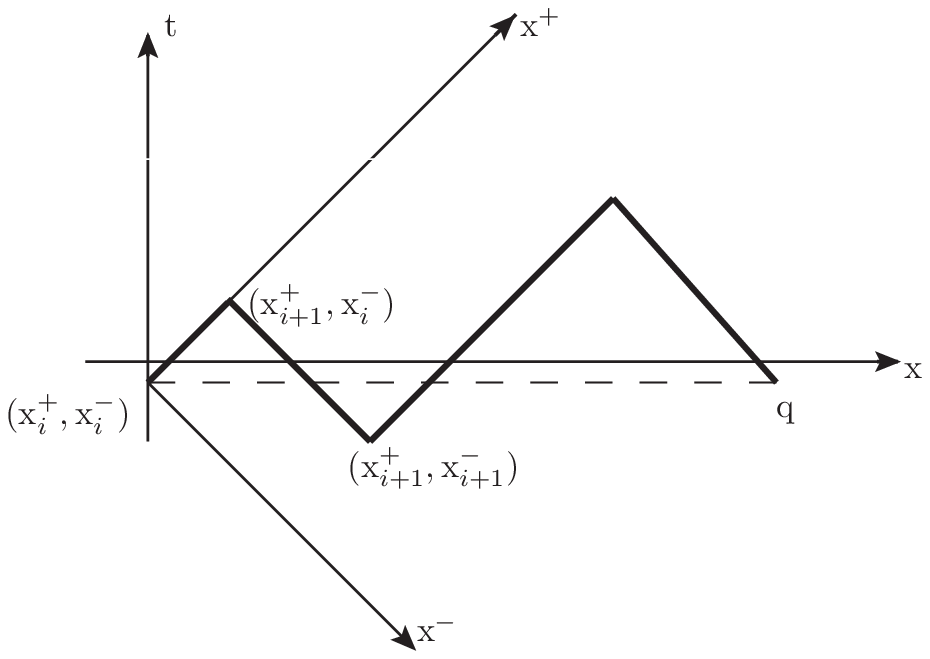}}

For the distance between two points on the boundary we use the following notation
\eqn\Lightcone{
x^{\pm}_{i j} = x^{\pm}_{i} - x^{\pm}_{j}
}
and we label the cusps as it shown at  \Label.

A point in   $AdS_{3}$ space can be conveniently written as an $SL(2,R)$ group element
\eqn\Embedding{ \eqalign{  Z_{a \dot a } =
\left(\matrix{Z_{-1}+Z_{2}& Z_1-Z_0\cr Z_1 + Z_0 & Z_{-1}-Z_{2} }\right)_{a \dot{a}}.
}}
The  $SO(2,2)$ symmetry of the model is realized by left and right multiplication by
$SL(2,R)\times SL(2,R)$.
We are interested in calculation of the divergent integral, we use the notation of \AldayYN ,
\eqn\AreaGen{
A= \int d x d y \pa_{i}Z \pa_{i} Z = 4 \int d^2 z e^{2 \alpha} = 2\int d^2 z {\rm Tr}[\Phi_{z} \Phi_{\bar{z}}].
}
The function $\alpha$ obeys  the generalized Sinh-Gordon equation \refs{\PohlmeyerNB,\DeVegaXC}
\eqn\SinhG{
\pa \bar{\pa} \alpha - e^{2 \alpha} + |p(z)|^2 e^{-2 \alpha} = 0.
}
Here $p(z)= \frac{1}{2} {\rm Tr}[\Phi_{z}^2]$ is a
holomorphic function in which the kinematics of the process is encoded.
This equation should be supplemented with appropriate boundary conditions for $\alpha$.

The problem of finding the a minimal area surface embedded in $AdS_{3}$ can be formulated
in terms of a ${\rm Z_{2}}$ projection of an $SU(2)$ Hitchin system where $\Phi$ is the Higgs field \refs{\HitchinVP}.
The ${\rm Z_{2}}$ projection acts on fields as follows $\Phi_{z} = - \sigma_3 \Phi_{z} \sigma_3$,
$\Phi_{\bar{z}} = - \sigma_3 \Phi_{\bar{z}} \sigma_3$, $A = \sigma_3 A \sigma_3$.

We study the flat sections, $\psi$,  that, by definition, obey the equation
\eqn\Section{
(d + {\cal A}) \psi = 0
}
with
\eqn\ConnectGen{
{\cal A} = {\Phi_{z} d z\over \zeta} + A + \zeta \Phi_{\bar{z}} d \bar{z}
}
here $\zeta$ is the spectral parameter and the original surface can be read off from
the solution of \Section\ at $\zeta=1,i$.

Recall that the worldsheet is parameterized by the whole complex plane $z$.
In the  large $z$ region some of the flat sections diverge. This means
that worldsheet goes to the boundary of $AdS$ space. The fact that the worldsheet  goes to different cusps
 is  due to  Stokes phenomena.

At large $z$ it is possible to diagonalize the Higgs fields
$\Phi(z) \sim \sqrt{p} \cdot {\rm diag}(1,-1)$, $\Phi(\bar{z}) \sim \sqrt{\bar{p}} \cdot {\rm diag}(1,-1)$.
This allows one to determine the form of two independent solutions of the section problem \Section\ at large $z$
\eqn\FlatSection{
\psi_{a} \sim \exp \left( (-1)^{a} \frac{\int dz \sqrt{p(z)}}{\zeta} +  (-1)^{a} \zeta \int d\bar{z} \sqrt{\bar{p}(\bar{z})} \right), \quad a=0,1
}
The number of Stokes sectors is determined by the number of cusps or equivalently by the degree
of polynomial. Within each Stokes sector the worldsheet reaches a different point at the boundary of $AdS$.
It is very useful to define the smallest solution $s_{i}$ in each of the
Stokes sectors. It is a flat section which has fastest decay rate as $z \to \infty$.
 This solution exists and it is unique in each Stokes sector,  up to an overall rescaling.

Importantly, in the case of an operator insertion we assume that the connection has a
 singularity   at $z=0$,  which corresponds to the operator insertion. The operator is specified
by the monodromy, $\Omega(\zeta)$,  of the connection around $z=0$.

We are going to study the problem at large $z$ with this monodromy which geometrically corresponds to certain periodic
Wilson loop. Importantly, the connection is single-valued and comes to itself when we go around the insertion point $z \to z e^{2 \pi i}$.
However due to the non-trivial monodromy,  the sections
 do not go to themselves   and the Wilson polygons are not closed.

Let us consider the small solution in the $i$'th Stokes sector (in some terminology these are anti-Stokes sector)
\eqn\StokesS{
(i - \frac{3}{2}) \frac{2 \pi}{n} + \frac{2}{n} {\rm Arg}[\zeta] < {\rm Arg}[z] < (i - \frac{1}{2}) \frac{2 \pi}{n} + \frac{2}{n} {\rm Arg}[\zeta]
}
which is denoted by $s_{i} (\zeta)$.
Due to $Z_2$ projection $\sigma_{3} s_{i} (e^{i \pi} \zeta)$ is also a solution of the
flat section problem. From the large $z$ form of the solution one can easily see that it is
a small solution in $(i+1)$'th Stokes sector. Thus, we can define $s_{i+1} (\zeta) = i \sigma_{3} s_{i} (e^{i \pi} \zeta)$ and normalize
$\la s_1 s_2 \ra = 1$. From the definition it follows that
\eqn\NormGen{\la s_{i} s_{j} \ra (e^{i \pi} \zeta) = \la s_{i+1} s_{j+1} \ra ( \zeta ).}
Taking into account the normalization we chose it leads to
\eqn\NormConcr{ \la s_{i} s_{i+1} \ra = 1.}
Throughout the paper we extensively use $SL(2)$ invariant inner product for flat sections defined as
\eqn\Wedge{
\la \psi_{a} \psi_{b} \ra = \psi_{a} \wedge \psi_{b} = \epsilon^{\alpha \beta} \psi_{\alpha,a} \psi_{\beta,b}.
}

\ifig\Label{The $z$ plane with Stokes sectors separated by black dotted lines. In each Stokes sector we have a unique small solution. Starting from $0$-th sector we have infinite chain $...,s_{-2},s_{-1},s_{0},s_{1},...$ of small solutions . To evaluate the inner product between the small solutions $s_{i}$ and $s_{j}$ with $i<j$ we need to specify a path of analytic continuation from the Stokes sector where small solutions are defined to the point where the inner product is evaluated. We choose it to be unfolded anti-clockwise path that connects sectors $i$ and $j$ which is given by solid blue line. The red dot corresponds to the operator insertion. The green dotted line indicates the one that we use to evaluate $\la s_{i} s_{j-n}\ra$. The blue dotted one serves to evaluate $\la s_{i} s_{j+n}\ra$. The wavy line allows one easily count the number of correspondent Stokes sector with correspondent small solution. In the case when there is no operator insertion $s_{i+n} \sim s_{i}$ and all paths are equivalent.
 }{\epsfxsize2.5in\epsfbox{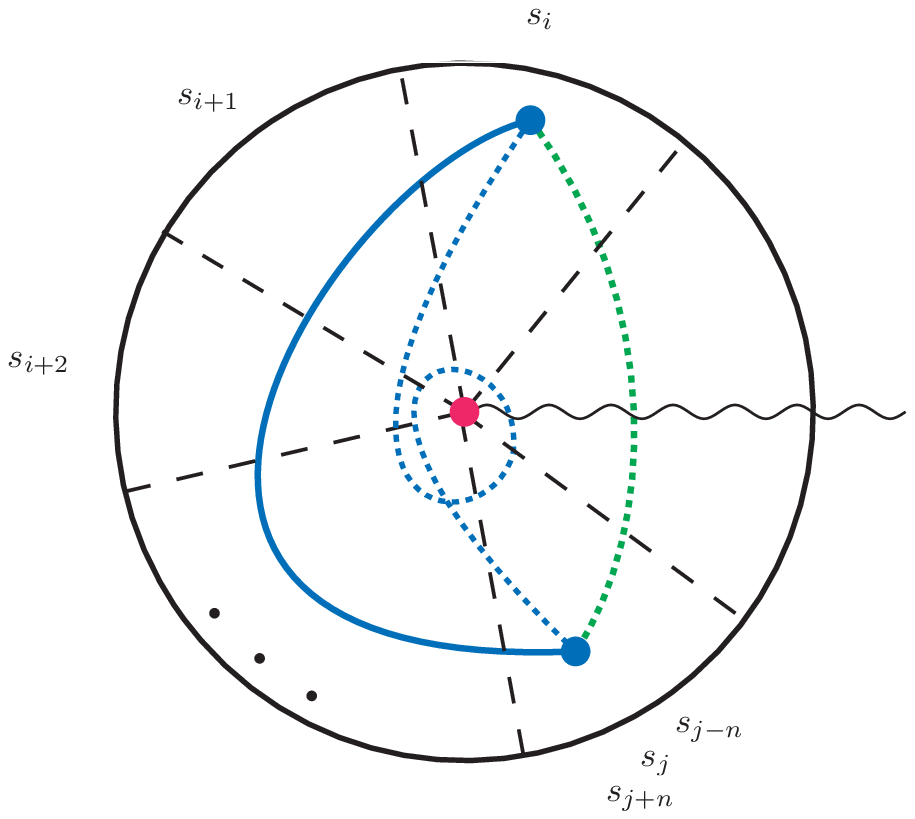}
 }

 \ifig\UNF{ We represent the same as in \Label\ but in the covering space, the $\tilde z $ plane. The two spaces are related through
 the map $z = e^{\tilde{z}}$. Here the boundary
 of the disk of \Label\ is to the right. The origin in \Label\ is to the left. The lines represent the same inner products
 as in \Label\ .
 }{\epsfxsize2.5in\epsfbox{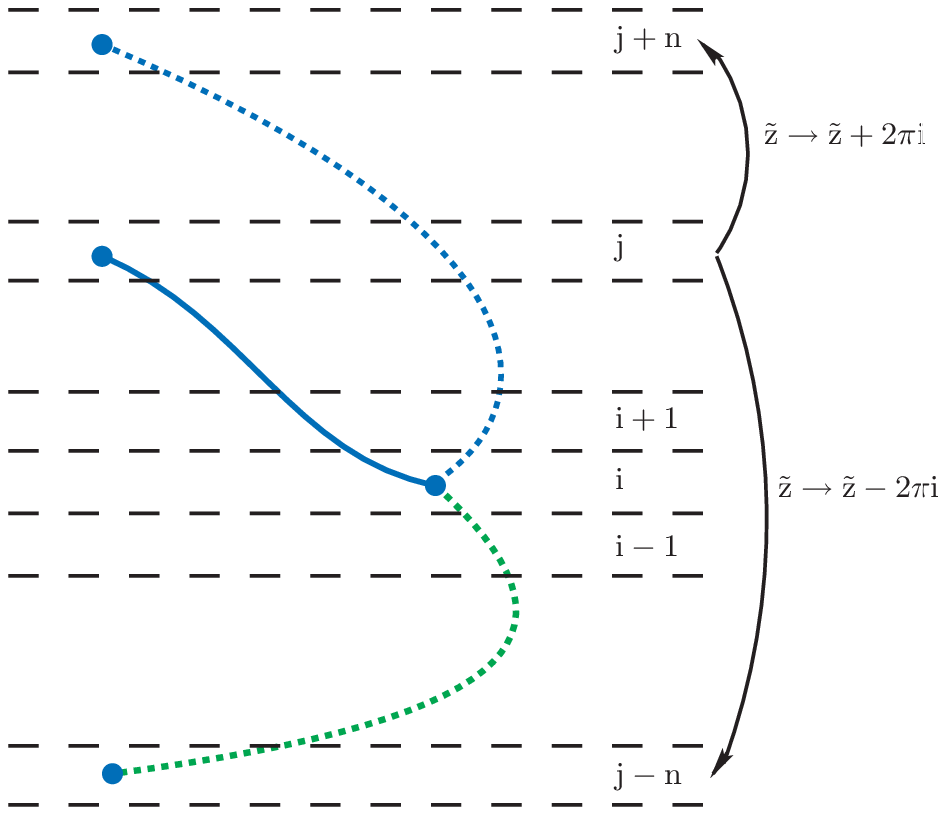}}
This product does not depend on the point $z$ where we evaluate it. However, it can depend on the path that
we choose to go from the region at infinity where $s_i$ are naturally defined to the point where we evaluate it.
In order to define the inner product in unique way we need to specify this path.
It is convenient to go to the covering space of the punctured disk, so that we think of $z$ and
$e^{ 2 \pi i } z $ as different points. The sections defined above are not periodic on the
punctured disk but they are naturally defined on the covering space. Note also that the sections
$s_i$ defined as the ones small in the sector \StokesS\ are really defined on the covering space. In
other words, the sectors in \StokesS\ are the sectors of the covering space. There is an infinite
number of such sectors. 
When we take inner products, such as \Wedge\ we need
to transport the sections to a common point. We do this transport in the covering space, which is
simply connected. Of course, once we project down to the disk, it might happen
that the path that we follow to evaluate an inner product might wind around $z=0$ a number of times.
This can be seen in \Label . In the covering space, the same figure looks as in \UNF .   So, for example, if we evaluate $\langle s_0 s_1\rangle $ we do not wind
around $z=0$. However, if we evaluate $\langle s_n s_0 \rangle$ then we wind once around $z=0$.
Note that $s_0$ and $s_n$ are not the same section. $s_n$ is the section that is small after we have
gone around $z=0$ once.

One can use the small solutions to define conformal cross ratios as functions of the spectral parameter as follows
\eqn\Cross{
\chi_{ijkl}(\zeta) = \frac{\la s_{i} s_{j} \ra \la s_{k} s_{l} \ra}{\la s_{i} s_{k} \ra \la s_{j} s_{l} \ra}.
}
Notice that these objects are independent of normalization of small solutions.
The kinematics of the process is then encoded in
\eqn\CrossLeftRight{
\chi_{ijkl}(1) = \frac{x_{i j}^{+} x_{k l}^{+}}{x_{i k}^{+} x_{j l}^{+}}~,~~~~~~\chi_{ijkl}(i) = \frac{x_{i j}^{-} x_{k l}^{-}}{x_{i k}^{-} x_{j l}^{-}}
}
We can derive functional equations for these objects as explained in \AldayVH.
We   use Schouten identity $$\la s_{i} s_{j} \ra \la s_{k} s_{l} \ra+\la s_{i} s_{l} \ra \la s_{j} s_{k} \ra+\la s_{i} s_{k} \ra \la s_{l} s_{j} \ra = 0$$ and the definition $T_{k} = \la s_{0} s_{k+1}\ra (e^{- i (k+1) \frac{\pi}{2}} \zeta)$ to get the Hirota equation
\eqn\Hirota{
T^{+}_{k} T^{+}_{k} = T_{k+1} T_{k-1} + 1
}
where $f^{\pm} = f(e^{\pm i \frac{\pi}{2}} \zeta)$ and \NormConcr\ were used.
Recall that from the definition and normalization convention it follows that $T_{-1}=0$ and $T_{0}=1$.
As the next step we introduce the Y-functions $Y_{s} (\zeta) = T_{s-1} T_{s+1}$ which contain all
the kinematic information
of the problem. They are  equal to conformal cross ratios at particular values of $\zeta$.
In terms of the Y-functions we get the set of equations
\eqn\Ysystem{
Y_{s}^{+} Y_{s}^{-} = (1+Y_{s+1})(1+Y_{s-1}) ~;~~~~~~~~~~~~~~ Y_0 =0 .
}
This set of equations is truncated from below because $Y_{0}$ is equal to zero according to
its definition.The equations will be  truncated from above by the periodicity condition, as we will explain below.
The precise condition   depends on $\Omega(\zeta)$.

\subsec{Kinematics of the general problem}

Let's count the number of parameters we have in the problem. On the
boundary of $AdS$ we have a polygonal Wilson loop which is periodic
up to the action of the two copies of $SL(2,R)$.

To be more precise we are thinking about the Wilson polygon with an infinite
number of cusps at the positions $(x_{i}^{+},x_{i}^{-})$. However the position of
the cusp $(x_{i+n}^{+},x_{i+n}^{-})$ is related to the position of $(x_{i}^{+},x_{i}^{-})$
through a conformal transformation
\eqn\Monodromy{
Z_{i+n} = \hat{\Omega}_{L} Z_{i} \hat{\Omega}_{R}.
}
Of course, these conformal transformations are the monodromies around the origin.
If we concentrate on $x^{+}$ then the Wilson loop is periodic up to the action of
$SL(2,R)$ transformation. Let us ask how many independent cross ratios we can make
with these constraints.

By conformal transformation we can fix the points $x_{i}^{+}$,~$i=0,1,2$.
Then we can have arbitrary points $x^{+}_{i}$,~$i=3,..,n+2$ where points
$i=n,n+1,n+2$ specify the monodromy. In other words, the monodromy is the unique
conformal transformation which maps the points $0,1,2$ to the points $n,n+1,n+2$.

Thus, including the degrees of freedom necessary to specify the monodromy
we have $2n$ variables.

This counting is just a property of the polygon at the boundary and does not depend on
the object inserted in the interior of the worldsheet.

\subsec{Truncation of the Y-system}

At the level of flat sections we introduce the monodromy matrix as follows
\eqn\DefMon{
\pmatrix{s_{1} \cr s_{0}} (z e^{2 \pi i}, \zeta) = \Omega(\zeta) \pmatrix{s_{1} \cr s_{0}} (z, \zeta)~,~~~~~~~ \langle s_0 s_1 \rangle =1
}
Recall that any two linearly independent solutions of the flat section problem form a basis and
here we choose $s_{0}$ and $s_{1}$ as basis solutions. $\Omega(\zeta)$ can be though as being known and
characterizing the operator.

By definition $s_{n}(z,\zeta)$ is the solution which is small in the same sector as $s_{0}(z,\zeta)$ but after
going around the complex plane once. So it should be proportional to $s_{0}(z e^{- 2 \pi i},\zeta)$
\eqn\Trunc{
s_{n}(z,\zeta)= B(\zeta) s_0 (z e^{- 2 \pi i},\zeta)
}
where we used the fact that the connection is single-valued.

Here $B(\zeta)$ appears due to the normalization convention and it
is obviously different from zero for any $\zeta$.
Changing the spectral parameter and multiplying by $i \sigma_3$ we get
\eqn\TruncTwo{s_{n+1}(z,\zeta)= B^{++}(\zeta) s_1 (z e^{- 2 \pi i},\zeta).}
If we take the wedge of \Trunc\ and \TruncTwo\ we get
\eqn\Bbb{B^{+}B^{-} = 1.}

Using these definitions we have
\eqn\TruncA{
\pmatrix{B s_{n+1} \cr B^{-1} s_{n}} (z,\zeta) = \pmatrix{s_{1} \cr s_{0}} (z e^{-2 \pi i}, \zeta) = \Omega^{-1}(\zeta) \pmatrix{s_{1} \cr s_{0}} (z, \zeta).
}
Taking the wedge we can calculate the trace of the monodromy
\eqn\TruncB{
B(\zeta) \la s_{0} s_{n+1} \ra(\zeta) - B^{-1}(\zeta) \la s_{1} s_{n}\ra(\zeta) = {\rm Tr} [\Omega^{-1}(\zeta)] = {\rm Tr} [\Omega(\zeta)]
}
here we used the fact that $\det[\Omega] = 1$ which immediately follows from the fact that the connection
\ConnectGen\ is traceless.
Using the definition for $T$ functions, \TruncB\  can be rewritten as
\eqn\TruncC{
B(\tilde{\zeta}) T_{n}(\zeta) - B^{-1}(\tilde{\zeta}) T_{n-2}(\zeta) = {\rm Tr} [\Omega (\tilde{\zeta})]
}
where $\tilde{\zeta} = e^{-i (n+1) \pi/2} \zeta$. Using this equation we can truncate the chain of Hirota equations
by expressing $T_{n}$ through $T_{n-2}$
\eqn\TruncD{
T_{n}(\zeta) = B^{-1}(\tilde{\zeta}) {\rm Tr} [\Omega (\tilde{\zeta})]+ B^{-2}(\tilde{\zeta}) T_{n-2}(\zeta)
= B^{-1}(\tilde \zeta) \left[ {\rm Tr}[ \tilde \Omega ] + \bar Y  \right]
}
where we introduced
\eqn\Ybar{
\bar{Y}(\zeta)  = B^{-1}(\tilde{\zeta}) T_{n-2}(\zeta) ~,~~~~~~~~~~~~ {\rm Tr} [\tilde \Omega ] = {\rm Tr} \Omega (\tilde \zeta).
}
Using $B^{+} B^{-} = 1$, we get the following Y-system for the general periodic Wilson loop
\eqn\Yoper{\eqalign{
Y_{s}^{+} Y_{s}^{-} &= (1 + Y_{s+1}) (1 + Y_{s-1}), \quad s=1,..., n - 3 \cr
Y_{n-2}^{+} Y_{n-2}^{-} &= (1 + Y_{n - 3}) (1  + {\rm Tr}[\tilde \Omega] \bar{Y} + \bar{Y}^2) \cr
\bar{Y}^{+} \bar{Y}^{-} &= 1 + Y_{n - 2}.
}}

Here the truncation condition was used to get the equation for $Y_{n-2}$.
We started with $Y_{n-2}^{+} Y_{n-2}^{-}=(1 + Y_{n - 3})(1 + Y_{n - 1})$ and then used
the fact that $Y_{n-1} = T_{n-2}T_{n} = T_{n-2} B^{-1} [ {\rm Tr}[\Omega(\tilde \zeta)]  + \bar Y ]  =
 {\rm Tr}[\tilde \Omega] \bar{Y} + \bar{Y}^2 $.

Note that due to the  ${\rm Z_{2}}$ projection the trace of the
monodromy ${\rm Tr}[\tilde \Omega] = {\rm Tr}[\Omega (\zeta)]$ for $n$ odd. For $n$ even, it is
simply ${\rm Tr}[\tilde \Omega] = {\rm Tr}[\Omega(\zeta)]^+$.

Of course, to choose the solution one needs to supplement these
equations with the analytic properties of Y-functions. Another issue is to relate the solution to
the expression for the area.

We immediately see that the number of cross ratios that appear in the system is $2(n-1)$.
Also we need to specify ${\rm Tr}[\tilde \Omega]$ at $(\zeta=1;~   i)$   to specify the polygon at the boundary.
Thus, including the monodromy we have $2 n$ parameters which agrees with the general counting above.

\subsec{Normalization independent definition of the Y-functions}

In the derivation above we use one particular normalization of small solutions.
It is convenient to write the $Y$ functions also for general normalizations.
 For $Y_s$ this is identical to the case of scattering amplitudes \refs{\AldayVH}
\eqn\GaugeInv{\eqalign{
Y_{2 k} &= \frac{\la s_{-k}, s_{k} \ra \la s_{-k-1}, s_{k+1} \ra}{\la s_{-k-1}, s_{-k} \ra \la s_{k}, s_{k+1} \ra} \cr
Y_{2 k + 1} &= \left(\frac{\la s_{-k-1}, s_{k} \ra \la s_{-k-2}, s_{k+1} \ra}{\la s_{-k-2}, s_{-k-1} \ra \la s_{k}, s_{k+1} \ra}\right)^{+}.
}}
Notice that to get physical values for cross ratios one needs to evaluate $Y_{2k} (\zeta=1,i)$ and $Y_{2k+1} (\zeta=1,i)$ correspondingly\foot{In general,  all $Y_s( \zeta = i^k)$ are physical cross ratios. The ones we mentioned are
a particular set of functionally independent ones. }.
One can think of points at the boundary of $AdS$ space as given
by points,  $\hat Z$, in $R^{2,2}$ with the constraint $\hat Z^2 =0$ and the identification
$\hat Z \sim \lambda \hat Z$. Then one can solve the constraint $\hat Z^2 =0$ introducing a pair
of spinors $\lambda $, $\tilde \lambda$, via $\hat Z_{a \dot b} = \lambda_a \tilde \lambda_{\dot b}$.
Where the $a$ index transforms under the left $SL(2)$ and the $\dot b$ index under the right $SL(2)$.
These are very simply related to $x^\pm$ coordinates, $x^+ = { \lambda_2 \over \lambda_1 }$,  $x^- = { \tilde
\lambda_2 \over \tilde \lambda_1 } $. As explained in \AldayVH\ we can
rewrite the cross ratios in terms of the positions of the cusps by replacing the $s_i$ in \GaugeInv\ by
$\lambda$ or $\tilde \lambda$. Thus the physical values of the Y-functions correspond to the
points of the unfolded infinite, periodic, polygons.

\ifig\YbarFig{ Inner products in the computation of $\bar Y$. We have two consecutive small sections on the disk. We take one
inner product going around the origin (in blue) and we divide by the same inner product but defined on the more direct path (in red).
 } {\epsfxsize1.0in\epsfbox{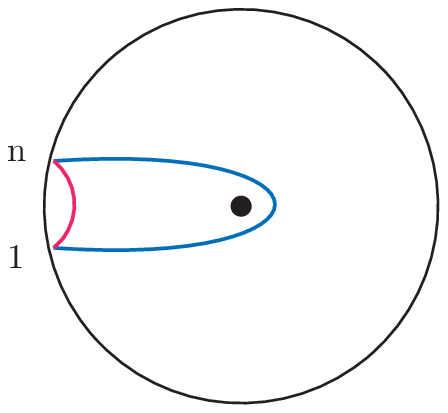}}

The new element in the case of the operator insertion is $\bar{Y}$.
We calculate $B$ by taking the inner product of \Trunc\ with $\tilde s_1 = s_1(ze^{-i 2 \pi })$.
We then get
\eqn\YbarA{\eqalign{
\bar{Y} (\zeta) &= \frac{\la s_{1} s_{n} \ra  }{\la s_{n} \tilde s_{1}\ra } (\zeta e^{- i \pi \frac{n+1}{2}}) =
- \frac{\la s_{1} s_{n} \ra  }{\la s_{1}  s_{n}\ra_{\rm Other~ Contour } } (\zeta e^{- i \pi \frac{n+1}{2}})
}}
Where the ``Other Contour'' is simply a contour which connects the two cups without going around the origin. It is the same
contour we would use to evaluate $\langle s_0 s_1 \rangle$.
A simple graphical representation of $\bar Y$ is in \YbarFig .

 The expression \YbarA\ defines $\bar Y$ in a normalization independent fashion, but
 it is not yet defined in terms of target space quantities that specify the shape of the polygon.
 In other words, we would like to derive an expression in terms of the $\lambda_i$ and $\tilde \lambda_i$ which specify
 the target space contour.
 This can easily be done. We can derive the expression
 \eqn\resfi{
  \bar Y ( \zeta = i^{ n+1}  ) = - { \la \lambda_1 ,\lambda_n \ra  \over \la  \hat \Omega \lambda_1 ,\lambda_n \ra }
  }
  where $\hat \Omega$ is the $SL(2,R)$ transformation that is uniquely determined by the condition that it (projectively)
  sends points $\lambda_0 , \lambda_1 , \lambda_2$ to $\lambda_n , \lambda_{n+1} , \lambda_{n+2}$. Then all
  the $\lambda_i$ obey, $\lambda_{n+i } \propto \hat \Omega \lambda_i $. We have a similar expression in terms of
   $\tilde \lambda_i$ and the value of $\bar Y$ at a shifted value of $\zeta$.

Alternatively, we can solve for $\hat \Omega$ in terms of the $\lambda$'s and write an expression which only
involves the target space positions of the points  (see Appendix G)
\eqn\YbarI{\eqalign{
\Lambda^2 &= { x_{ {n}+ {2},  {n}+ {1}} \over x_{ {2},  {1}}} { x_{ {2},  {0}} \over x_{ {n}+ {2},  {n}} } { x_{ {n}+
{1},  {n}} \over x_{ {1},  {0}}}, \cr
\bar{Y}(\zeta = i^{n+1}) &= { x^{+}_{ {n},  {1}} \over x^{+}_{ {n}+ {1},  {n}}}  \Lambda^{+}, \cr
\bar{Y}(\zeta = i^{n+2}) &= { x^{-}_{  {n},  {1}} \over x^{-}_{ {n}+ {1},  {n}}}  \Lambda^{-}.
}}
Of course, we can also express this in terms of the $\lambda_i$ by replacing
$x^+_{i,j} \to \la \lambda_i \lambda_j \ra$, $ x^-_{i,j} \to \la \tilde \lambda_i \tilde \lambda_j \ra $ .

\newsec{Constant monodromies}

The functional equations \Yoper\ are valid for the insertion of a generic operator. If it is an operator that can
be described by a semiclassical string state, then we should put in \Yoper\ the monodromy of the corresponding
state. In fact, for the so called ``finite gap'' states the monodromy can be computed in terms of an auxiliary
hyper-elliptic Riemann surface. We will not derive the full formula for the area in these cases in this paper.

More generally, the equations \Yoper\ are valid even if a more complicated object is inserted in the $z \sim 0$ region.
Of course, those more complicated objects would have more intricate monodromies, about which we know little. The full
computation of the area will, almost certainly, require the computation of extra Y-functions beyond the ones we have
introduced.

In his section, we   concentrate on the simplest case. We consider   monodromies  independent of the spectral parameter.
We describe below some special cases, one of which corresponds to the insertion of operators with small conformal dimension.
In this cases we will later derive   the full formula for the area.

\subsec{Recovering the Y-system for scattering amplitudes}

Let us  consider the simplest possible monodromy
\eqn\Amplit{\eqalign{
\hat{\Omega} &= \left( \matrix{1 & 0 \cr 0 & 1} \right) \cr
{\rm Tr}[\hat{\Omega}] &= {\rm Tr}[\Omega] = 2
}}
In this case, as we go around  $z=0$
the solution goes to itself. All sections are single valued and
 $  s_{i} \sim s_{i+n}$ and $x_{i+n} = x_{i}$.
We see that $T_{n-1} = \la s_{0} s_{n}\ra = 0$ and so $Y_{n-2} = 0$.
From the definition \YbarA\ we find $\bar{Y}=-1$.
Thus, the Y-system is reduced to the on in \AldayVH
\eqn\Yamp{
Y_{s}^{+} Y_{s}^{-} = (1 + Y_{s+1}) (1 + Y_{s-1}), \quad s=1,..., n - 3
}

\subsec{Y-system for the form factors of operators with small dimension}

The next case we would like to consider is
\eqn\OpIns{\eqalign{
\hat{\Omega} &= \pmatrix{1 & 0 \cr q & 1} \cr
{\rm Tr}[\hat{\Omega}] &= {\rm Tr}[\Omega] = 2.
}}

Geometrically it corresponds to a solution which as we move around the operator insertion $z \to z e^{2 \pi i}$
 transforms as $x(z e^{2 \pi i}) = x(z) + q$,  $t(z e^{2 \pi i}) = t(z)$. It means that we consider solutions
 which are periodic in $x$ direction
with period  $q$. For the cusp positions it means $x^{\pm}_{i+n} = x^{\pm}_{i} + q$. As reviewed above, these solutions
 are the strong coupling dual of  form factors of
operators with small dimension.   $q$ is  momentum carried by the operator.

From the formula \YbarI\ it follows that (for this case $\Lambda =1$)
\eqn\YbarIns{\eqalign{
\bar{Y}(1) &=
\frac{q - x_{-\frac{n-1}{2},-\frac{n+1}{2}}^{+}}{x_{-\frac{n-1}{2},-\frac{n+1}{2}}^{+}}, \cr
\bar{Y}(i) &=
\frac{q - x_{-\frac{n-1}{2},-\frac{n+1}{2}}^{-}}{x_{-\frac{n-1}{2},-\frac{n+1}{2}}^{-}}
}}
Thus, we see that $\bar{Y}$ is a conformal invariant way of defining the period of the form factor solution.
The Y-system in this case takes the form
\eqn\Yoperins{\eqalign{
Y_{s}^{+} Y_{s}^{-} &= (1 + Y_{s+1}) (1 + Y_{s-1}), \quad s=1,..., n - 3 \cr
Y_{n-2}^{+} Y_{n-2}^{-} &= (1 + Y_{n - 3}) (1 + \bar{Y})^2 \cr
\bar{Y}^{+} \bar{Y}^{-} &= 1 + Y_{n - 2}.
}}

Since  ${\rm Tr}[\Omega]=2$ but $\bar{Y}\neq -1$ the number of free parameters in the
problem is $2(n-1)$.

This kinematical counting can be understood as follows.
 In this case we need $2(n+1)$ coordinates to fix the momenta of $2n$ gluons.
 Restricting the monodromy to be a pure translation (rather than a more general conformal
 transformation) implies that we get
  $2(n-1)$ parameters. Alternatively, we can say we have $ 2n $ light-like momenta, with $ 2n$ parameters.
  Boosts and dilatations allow us to reduce this to $2(n-1)$ parameters.

\subsec{Y-system for ${\rm Z_{m}}$ symmetric polygons}

As the last example we would like to consider the following monodromy
\eqn\Zm{\eqalign{
\hat{\Omega} &= \pmatrix{\cos( \frac{\pi}{m} )& - \sin( \frac{\pi}{m} ) \cr \sin( \frac{\pi}{m} )& \cos(\frac{\pi}{m})} \cr
{\rm Tr}[\hat{\Omega}] &= {\rm Tr}[\Omega] = 2 \cos(\frac{\pi}{m})
}}
We should emphasize that it is again $\zeta$ independent.

This corresponds to  solutions where the global $AdS_3$ boundary coordinates  obey
$\phi(z e^{2 \pi i}) = \phi(z) + \frac{2 \pi}{m}$, $\tau(z e^{2 \pi i}) = \tau(z)$.
Thus, we are dealing with ${\rm Z_{m}}$ symmetric solutions. From the physical point of view they correspond to calculation of the scattering amplitudes
with ${\rm Z_{m}}$ symmetric kinematics.
In the case when number of cusps is equal to $2m$ these are regular polygons and they were
considered in the literature before. More generally these are special subcases of Wilson loops with $ 2n m$ cusps.

From the formula \YbarI\ it follows that
\eqn\YbarZM{\eqalign{
\delta \phi &= \phi_{-\frac{n-1}{2},-\frac{n+1}{2}} \cr
\bar{Y}(1) &= \frac{\sin({\pi \over m} - {\delta \phi^{+} \over 2})}{\sin({\delta \phi^{+} \over 2})},\quad \phi^{+}=\phi + \tau \cr
\bar{Y}(i) &= \frac{\sin({\pi \over m} - {\delta \phi^{-} \over 2})}{\sin({\delta \phi^{-} \over 2})},\quad \phi^{-}=\phi - \tau
}}

The Y-system takes the following form in this case
\eqn\YZM{\eqalign{
Y_{s}^{+} Y_{s}^{-} &= (1 + Y_{s+1}) (1 + Y_{s-1}), \quad s=1,..., n - 3 \cr
Y_{n-2}^{+} Y_{n-2}^{-} &= (1 + Y_{n - 3})(1 + e^{ \frac{\pi}{m} i } \bar{Y})(1 + e^{ - \frac{\pi}{m} i }\bar{Y}) \cr
\bar{Y}^{+} \bar{Y}^{-} &= 1 + Y_{n - 2}.
}}

We can do one check of these equations immediately using the solution for the simplest case of regular polygons in $AdS_{3}$ which
was considered in the literature \refs{\Kirillov}
\eqn\YZMb{
Y_{s} = \frac{\sin(\frac{\pi (s+2)}{n m})\sin(\frac{\pi s}{n m})}{\sin^2 (\frac{\pi}{n m})}
}
here the total number of gluons is $2 n m$ and the symmetry of the problem is $Z_{n m}$.
However, we would like to consider this polygon as being ${\rm Z_{m}}$ symmetric with $2n$ cusps over one period and apply the Y-system written above to it.

From the \YbarZM\ we get plugging $\delta \phi^{\pm} = {2 \pi \over n m}$ that
\eqn\YZMc{
\bar{Y} = \frac{\sin(\frac{\pi (n-1)}{n m})}{\sin(\frac{\pi}{n m})}.
}
One can check that with these expressions the equations \YZM\ are indeed satisfied.

These considerations allow us to make a  connection to the family of solutions of modified Sinh-Gordon recently considered
by Lukyanov and Zamolodchikov \refs{\LukyanovRN}.
We relate the parameter $l$ in \LukyanovRN\ to
 our $m$ via  $l = \frac{1}{2m} - \frac{1}{2}$. Then the solutions described in \LukyanovRN\ are special cases
 of the ones we consider here.
When $m=1$ we come back to the problem of amplitudes.
 When $m=\infty$ we end up with the problem of form factor.
  For any integer $m$ we
are dealing with the problem of calculation of the area of ${\rm Z_{m}}$ symmetric polygons.
This interplay is described in more detail in Appendix B.
 We do not know the  physical interpretation for these solutions at general values of $l$.

\newsec{Y-system for the form factors}

In this section we focus our attention on the problem of minimal area surfaces in the
 massless BTZ black hole
background,  which corresponds to the problem of the form factor calculation. Firstly, we consider zig-zag
solution to get the basic picture behind the problem. Then we turn to the analysis of analytical properties
of Y-functions which allows us to write the integral form of the equations.

\subsec{Zig-zag solution}

\ifig\Zigzag{The qualitative form of zig-zag solution which corresponds to the Sudakov form-factor. Here the Wilson line corresponds to momenta of gluons and
is situated at $r=0$ operator is inserted at $r = \infty$. $\la  k_{2} | O(q) |k_{1} \ra$ is related to the area of one period.
 } {\epsfxsize2.5in\epsfbox{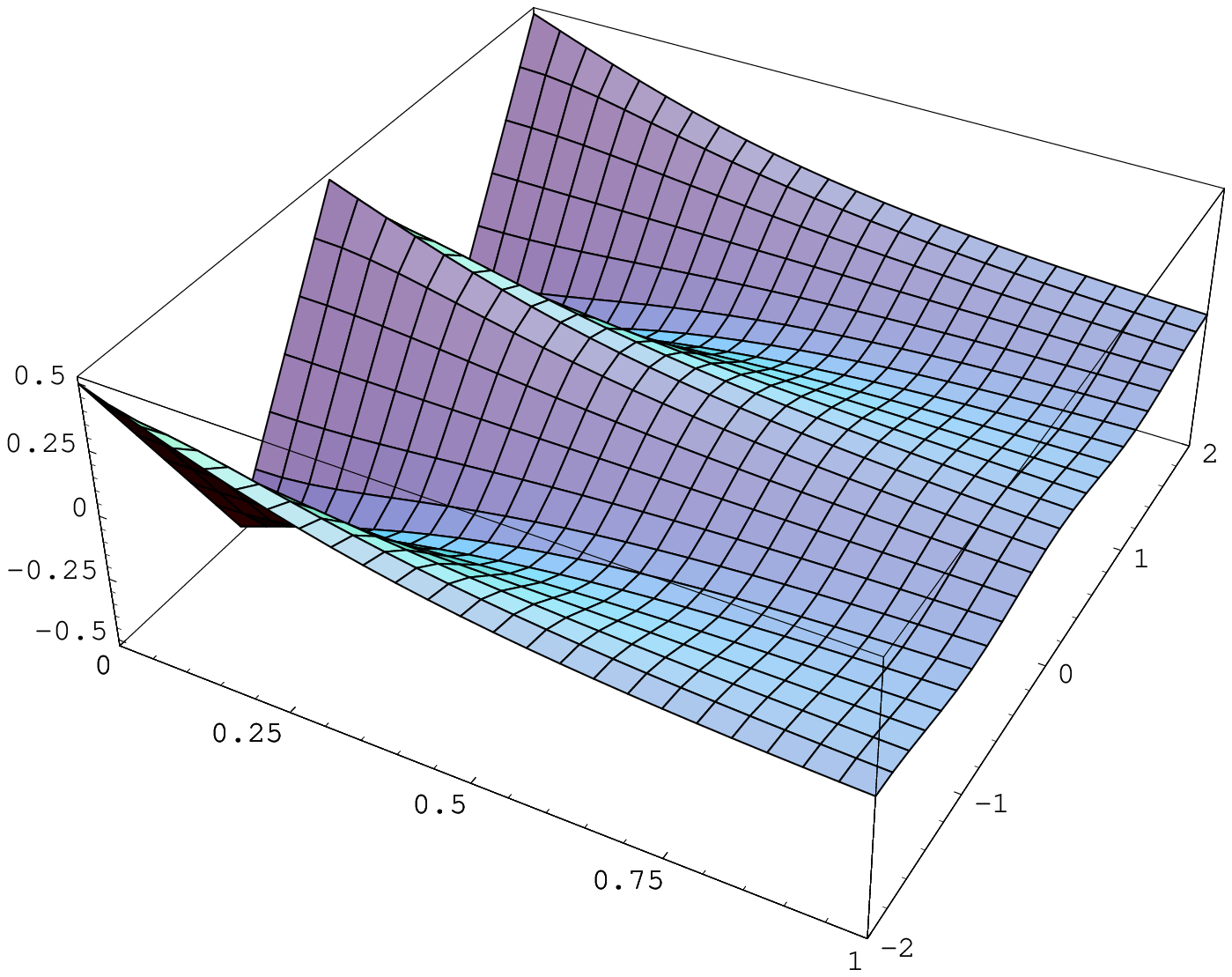}}

Here we analyze a  zig-zag spacelike classical solutions
in $AdS_3$ which corresponds to the following object at leading $\sqrt{\lambda}$
order:
\eqn\FormB{\eqalign{
\la k_{2} | O(q) |k_{1}\ra \cr
q = - ( k_1 + k_2 )
}}
This is an operator decaying into two gluons.
We are working in Poincare patch with the metric
\eqn\AdSPoinc{
ds^2 = \frac{dr^2 - dt^2 + dx^2}{r^2}
}
We parameterize the worldsheet as $t(x,r)$ and introduce $\tilde{z}= r + i x$. For large
$\tilde z$, or large $r$ the solution approaches the straight line Wilson loop solution which has $t=0$.
In that case,   the  induced metric  takes the form
\eqn\Induced{
ds^2_{ind} = \frac{d\tilde{z} d\bar{\tilde{z}}}{(\tilde{z} + \bar{\tilde{z}})^2}.
}
and the   $p(\tilde{z}) = 0$.
In order to implement the periodicity
we can make the standard map from the strip to the unit disc
\eqn\Map{
z = e^{-\tilde{z}}
}
then we have that near the origin induced metric takes the form
\eqn\IndMap{
ds^2_{ind} = \frac{dz d\bar{z}}{z \bar{z} \log^2(z \bar{z})} = e^{- 2  \alpha} dz d\bar{z}.
}
We will want to consider solutions which approach \IndMap\ when $z \to 0$.
In particular, note  that if we   consider the polynomial $p(z)= {1 \over z}$ then in the $\tilde{z}$ coordinates
it corresponds to $\tilde{p}(\tilde{z}) = ({\pa \tilde{z} \over \pa z})^{-2} p(z) = e^{-\tilde{z}}$.
For large $\tilde{z}$ the polynomial goes to $0$ as we expect it for the class of solutions we consider.
So we conclude that $p(z)={1 \over z}$ is allowed in the class of solutions we want to consider.
 Notice that for higher poles we get $\tilde{p}(\tilde{z})$
that does not go to zero at large $\tilde{z}$, and we would change the qualitative form of the solution near $z \sim 0$.

It is natural then to expect that the form of the polynomial for $2n$ gluons as the form
\eqn\Polynom{
p(z)= a_{n-2} z^{n-2}+a_{n-3} z^{n-3}+ a_{n-4} z^{n-4}+... + a_{0} + \frac{1}{z}
}
where we used translations to set the pole at $z=0$ and rescaling to set the
coefficient in front of the pole to one\foot{Recall
 that the pole in the polynomial we consider is not what was called a ``pole'' in \refs{\GaiottoHG}.
 What was called a ``pole'' in their construction corresponds to the double pole in the
  polynomial we consider.}. We see that we have $n-1$ complex parameters $a_i$.

Let us  summarize the picture that arises for the modified Sinh-Gordon in the table\foot{Solutions with given properties appeared in the past \refs{\McCoyCD}.}

\vskip3em
\vbox{
\begintable
{\rm Case of } $2n$ {\rm gluons}  | {\rm Scattering Amp.}  | {\rm Operator Insertion}
  \cr
  {\rm Polynomial}   |$p(z) = a_{0} + a_{1} z + ... + z^{n-2} $| $p(z) = \frac{1}{z} + ... + a_{n-2} z^{n-2}$
  \cr
  {\rm BC at $z = 0$}   | $\alpha$ {\rm regular}  | $\alpha \sim - \frac{1}{2} \log(z \bar{z} \log^2(z \bar{z}))$
 \cr
  {\rm BC at $z = \infty$}   | $\hat{\alpha} \to 0$ | $\hat{\alpha} \to 0$
  \endtable}
\centerline{{ Table~1:\/} Boundary conditions and polynomials in Modified Sinh-Gordon}
\centerline{for operator insertion and scattering amplitudes.}
\medskip

We see that the kinematic information for $2n$ gluons plus operator insertion is encoded in $2(n-1)$ real parameters of the polynomial which
is what we expected from the Y-system counting. In the Appendix A we show that given boundary conditions for $\alpha$, \IndMap\
 and the general form of the polynomial \Polynom\
 indeed lead to the off diagonal spectral independent monodromy which we used before.

Using this prescription and the techniques  of \refs{\ZamolodchikovUW, \AldayYN} we can find the area of the
one period for the zig-zag solution
\eqn\SinhZZ{
A_{Sinh}^{zz} = 4 \int d^{2} w ( e^{2 \hat{\alpha}} - 1) =  \frac{3 \pi}{4}
}
This is the answer for $n=1$, where we have no kinematic parameters.
The simplest way to get this
 answer is to consider the limit $n \to \infty$ of the area for regular polygons and
notice that $A_{Sinh}^{zz} = \lim_{n \to \infty} \frac{A_{Sinh}^{reg}(n)}{n} = \frac{3 \pi}{4}$, where we used the formula
for the area of the regular polygon $A_{Sihn}(n)$ (formula  (4.9) in \AldayYN ).

\subsec{Analytic properties of the Y-functions}

From the definition of $T$'s it is clear that they are analytic functions of
$\zeta$, for $\zeta \neq 0, \infty$. From this fact and the choice of normalization we made
it also follows that $Y_{s}$ for $s \leq n-2$ are analytic away from $\zeta = 0, \infty$ where they have
an essential singularity. For $\bar{Y}$ it follows from    \Ybar\ and the fact that $B\neq 0$ for any $\zeta$.

For $\zeta \to 0, \infty$ we can analyze the flat section problem using a WKB analysis with $\zeta$ playing the role of
$\hbar$. It was   developed in previous work on the subject \GaiottoHG . It was applied to the particular case
of scattering amplitudes in \AldayVH . Here
the only new ingredient is
the pole in the polynomial.

To be concrete let us  consider the case when $\zeta \to 0$. Remember that we are considering the equation
\eqn\Small{
(d + A + \frac{\Phi_{z} d z}{\zeta} + \zeta \Phi_{\bar{z}} d \bar{z}) s = 0.
}

For $\zeta \to 0$ it is convenient to make a complex gauge transformation which diagonalizes the Higgs field
\eqn\Higgs{
\Phi_{z} = \left( \matrix{\sqrt{p} & 0 \cr 0 & - \sqrt{p}} \right).
}
Intuitively it is clear that the solutions of the flat section problem are dominated by the term $\frac{1}{\zeta} \Phi_{z}$ and
have the approximate form $\left( \matrix{ e^{-\frac{1}{\zeta}\int \sqrt{p} d z} \cr 0} \right)$ or $\left( \matrix{ 0 \cr e^{\frac{1}{\zeta}\int \sqrt{p} d z}} \right)$.

To make this statement more precise it is useful to introduce several notions.
First of all it is natural to think about the problem as living on the
Riemann surface $\Sigma$ given by
\eqn\RS{
x^2 = p(z).
}
On the complex plane ${\cal C}$ it corresponds to introducing
branch cuts and working on two sheets.
 The differential $\lambda = \sqrt{p} dz$ that plays an important role
is also living on the $\Sigma$ so on the complex plane and we should be careful about the sheet
we are working on, namely $\lambda_i = (\sqrt{p}, - \sqrt{p})$ where $i$ is the sheet label.

Let us introduce the notion of WKB line, as a line where the exponent varies most rapidly
\eqn\WKB{
{\rm Im}[\frac{\dot{z} \sqrt{p(z)}}{\zeta}] = 0.
}
WKB lines live on the complex plane ${\cal C}$. Through each point on the complex plane only one WKB line is going through.
The direction of the WKB line shows the direction towards which the solution increases.

Then more precise statement is that small solutions in the limit $\zeta \to 0$ take the form
\eqn\SmallS{\eqalign{
s \sim c(z) \left( \matrix{ e^{- \frac{1}{\zeta} \int \lambda_1} \cr 0} \right) \cr
s \sim c(z) \left( \matrix{ 0 \cr e^{- \frac{1}{\zeta} \int \lambda_2}} \right).
}}
where $c(z)$ does not depend on $\zeta$. Thus, $s_{i}$ lives on the $(i+1)$-th sheet ${\rm mod}(2)$.

An important observation is that if we have a WKB line which connects Stokes sectors $i$ and $j$,
 we can use it to evaluate $\la s_{i} s_{j} \ra$.
Using this fact we can reliably find the asymptotic behavior of the Y-functions \GaiottoHG .

\subsec{Evaluating Y's}

\ifig\WKBlines{Here we present the approximate form of the flow for $\zeta = e^{i \frac{\pi}{4}}$ and $n=7$. Black dots
represent zeroes of $p(z)$ while the red one is the pole. The numbers label different Stokes sectors.
Blue and red lines which ends on zeroes separate different groups of WKB lines. The orange lines show the cycle along
which we evaluate $Y_{1}$ (on the right) and $\bar{Y}$ (on the left).}
{\epsfxsize4.0in\epsfbox{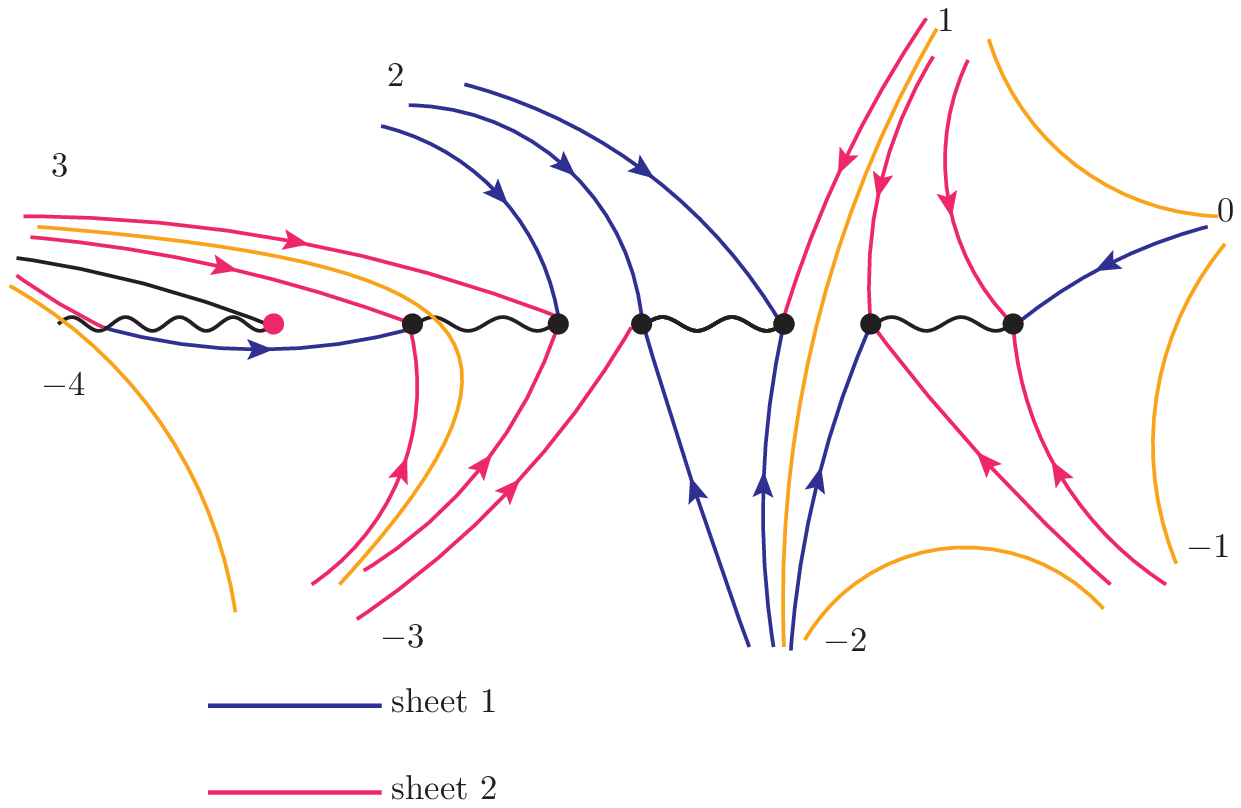}}

For convenience we choose $p(z) = \frac{\prod ({z \over z_{i}} - 1)}{z}$ with all zeroes being positive and located on real axis.
The only difference with the current situation compared to the amplitude case   is  the different flow pattern around the
pole, which we located to the left, see \WKBlines  . There are three lines ending at each zero and one line ending at the pole.
The lines that end at the zeros or poles separate different flow patterns of WKB lines (or different basins of attraction).

The asymptotic behavior of the
Y-functions at $\zeta=0,\infty$ is again given by contour integrals around   certain cycles.
Compared to the amplitude case, for the same value of $n$,  added an extra pole and an extra zero to the polynomial.
Thus, two more
  cycles appear. This is another manifestation of appearance of $2$ additional Y-functions in the
Y-system for the form factor compare to the case of amplitudes.

The only new feature here is the evaluation of the $\bar Y$ function. Form the definition,
\YbarA\ and figure \YbarFig , it is clear that we can evaluate it using the contour
indicated in figure \WKBlines .

The story happens to be completely identical to the amplitudes one. Here we just present the results.
Asymptotic behavior of Y-functions is given by (introducing $\zeta = e^{\theta}$)
\eqn\Ywkb{\eqalign{
\log Y_{s} &= - m_{s} \cosh \theta + ..., \quad s= 1,..., n-2 \cr
\log \bar{Y} &= - \bar{m} \cosh \theta + ...,
}}
where
\eqn\Ymasses{\eqalign{
m_{2k} &= 2 \oint_{\gamma_{2k}} \lambda \cr
m_{2k+1} &=  -2 i \oint_{\gamma_{2k+1}} \lambda \cr
\bar{m} &= 2 \oint_{\bar{\gamma}} \lambda
}}
these formulas are valid because with our choice of the polynomial
all masses are real and positive. Here one should be as usually careful
with the contour orientation and the sheet where the differential is considered.

If we slightly move  the  zeroes   $p(z)$  away from the real axis the $m_s$ in \Ymasses\ become complex.
The asymptotic behavior of Y-functions  is $\log Y_s \sim - { m_s \over2 }  e^{- \theta} $ for $\theta \to + \infty$ and
$\log Y_s \sim - { m_s^* \over 2 }  e^{\theta} $ for $\theta \to - \infty$.

\ifig\Cycles{The cycles which we need to integrate $\lambda$ along to determine the asymptotic behavior
of all Y-functions (case $n=7$). For the general odd $n$ intersection form is $\theta^{s r} = (-1)^{s+1} (\delta_{s+1,r} + \delta_{s-1,r})$.}
{\epsfxsize4.0in\epsfbox{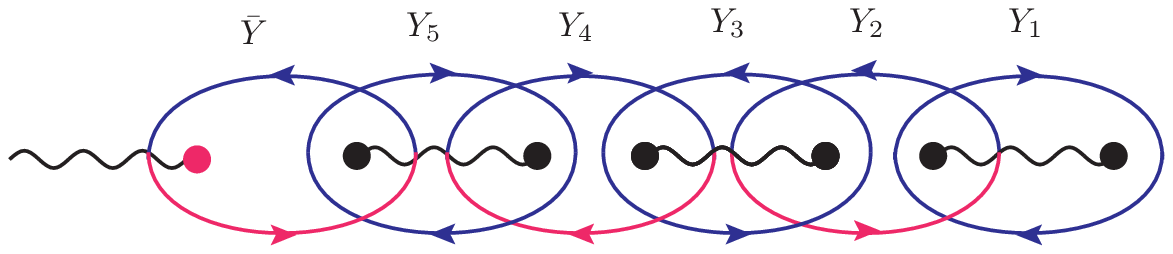}}

\subsec{Integral form of the equations}

Functional equations supplemented with the asymptotic behavior of the Y-functions specifies
the solution of Y-system uniquely. It is convenient to rewrite the equations in an integral form
which is completely analogous to the case of amplitudes.
If we introduce the kernel $K(\theta) = \frac{1}{2 \pi \cosh \theta}$ and the operation star being the convolution $K \star f = \int_{-\infty}^{\infty} d \theta' K(\theta - \theta') f(\theta')$ then the system of integral equations for the Y-functions as functions
of the spectral parameter takes the form
%
\eqn\Yintegral{\eqalign{
\log Y_{s} &= - m_{s} \cosh \theta + K \star \log (1 + Y_{s+1})
+ K \star \log (1 + Y_{s-1}), \quad s=1,..., n - 3 \cr
\log Y_{n-2} &= - m_{n-2} \cosh \theta + K \star \log (1 + Y_{n - 3})
+ 2 K \star \log (1 + \bar{Y})  \cr
\log \bar{Y} &= - \bar{m} \cosh \theta + K \star \log (1 + Y_{n - 2})
}}
where the fact that masses are real was used. The form of the integral equations in the general case of complex masses
is very similar and it  can be found in Appendix D.
If the phases of
the $m_s$ become very large, then one needs to perform wall crossing transformations.

We can also consider the case when the trace of the monodromy is ${\rm Tr} [\tilde \Omega ]= 2 \cosh ( {\mu \over \zeta} + \bar \mu \zeta )$.
This could arise if we replace the single pole at $z=0$ by a double pole and a closely located zero, $1/z \to {\mu^2 \over
4 \pi^2 \epsilon }  { z - \epsilon
\over z^2 }$ and also claim that $\hat{\alpha} \to 0$ near the origin\foot{These boundary conditions are natural from the \refs{\GaiottoHG} point of view.}.
In this situation, integral  equations continue to be the same as long as $2 |\mu | < |\bar m|$.
When this condition is not obeyed, it might be necessary to do wall crossing.
Note that in this case we can define new $Y$ functions $ \bar Y_{\hat \pm } = e^{ \mp ({\mu \over \zeta} + \bar \mu \zeta ) }\bar Y$.
We can then write the two final equations of the Y systems as
\eqn\newf{ \eqalign{
Y_{n-2}^+ Y^-_{n-2} & = (1 + Y_{n-3} ) ( 1 + \bar Y_{\hat +}) (1 + \bar Y_{\hat -} )
\cr
\bar Y_{\hat +}^+ \bar Y_{\hat +}^- & = (1 + Y_{n-2} ) ~,~~~~~~~~~~~~~\bar Y_{\hat - }^+ \bar Y_{\hat - }^-   = (1 + Y_{n-2} )
}}
The masses then are $\bar m_{\hat \pm } = \pm 2 \mu + \bar m $.

\newsec{Area in the case of form factors }

In this section we present the formula for the area first in the form
which contain mass parameters and then in the form which depends only
on cross ratios. We consider the case that the monodromy is constant and off diagonal, which corresponds to
the computation of form factors of operators with small conformal dimensions.

\subsec{General formula}

We start from the following divergent integral\foot{To be more precise the area $4\int d^{2}z e^{2 \alpha} = 2\int d^2 z {\rm Tr}[\Phi_{z} \Phi_{\bar{z}}] + 2 \int d^2 z \partial \bar{\partial}\alpha$.
For the case of form factor $2 \int d^2 z \partial \bar{\partial}\alpha = \frac{\pi n}{2}$. A term proportional to $n$ can be absorbed by a redefinition of the regularization procedure of the cusps. } to calculate
\eqn\Area{
A = 2\int d^2 z {\rm Tr}[\Phi_{z} \Phi_{\bar{z}}]
}
one can rewrite it as follows (for $n$ odd)
\eqn\ArPieces{\eqalign{
A &= A_{reg} + A_{periods} + A_{cutoff} \cr
A_{reg} &= A_{Sinh} = 4 \int d^{2}w (e^{2 \hat{\alpha}} - 1)=\int d^{2}z (2 {\rm Tr}[\Phi_{z} \Phi_{\bar{z}}] - 4 (p \bar{p})^{1/2})\cr
A_{periods} &= 4 \int d^{2}z (p \bar{p})^{1/2} - 4 \int_{\Sigma_0} d^2 w = 4 \int_{\Sigma} d^2 w - 4 \int_{\Sigma_0} d^2 w \cr
A_{cutoff} &= 4 \int_{\Sigma_0, z_{AdS}> \epsilon} d^2 w.
}}

The simplest part is $A_{cutoff} = A_{div} + A_{BDS-like}$

\eqn\ADivBds{\eqalign{
A_{div} &= \sum_{i=1}^{2n} \frac{1}{8} (\log \epsilon^2 s_{i,i+2})^2 \cr
A_{BDS-like} &= l_{i}^{+} M_{i j} l_{j}^{-} \cr
l_{i}^{+} &= \log (x_{i+1}^{+} - x_{i}^{+}), \quad l_{i}^{-} = \log (x_{i+1}^{-} - x_{i}^{-})
}}
where again $l^{\pm}_{i+n}=l^{\pm}_{i}$ and $M_{ij}$ can be read off from $(5.8)$ in \refs{\AldayDV}. The proof that it takes the same form as
in the case of amplitudes is given in the Appendix C. Qualitatively it can be understood from the fact that both terms
depend only on the difference of coordinates between two consecutive cusps so at most the part which involves the difference between
the $n$-th and first cusp can change due to the presence of the monodromy.
However, in the case of periodic polygons even this does not happen.

\subsec{Area as the free energy}

The non-trivial finite piece of the area is

\eqn\Asinh{
A_{Sinh}^{ff} = A_{free}^{ff} + C_{zz} + (n-1) C_{6} ~,~~~~~~~C_{zz} = { 3 \pi \over 4} ~,~~~~~~~~C_{6} = { 7 \pi \over 12 }
}
here $C_{zz}  $ is the area of the zig-zag solution \SinhZZ\ and $C_{6}  $ is the area of the regular hexagon.
One can get this formula as follows. When all zeroes and the  pole are far from each other the
free energy for the form factor Y-system goes to zero. Since  $\hat{\alpha}$
is massive field we get isolated contributions from the zeroes and the pole. Each pole correspond to the
regular hexagon while the pole to the area of the zig-zag solution.

For $A_{periods}$ one gets the same expression as in the case of amplitudes \refs{\AldayVH},
namely
\eqn\Aperiods{
A_{periods} = - i w_{s, s'} Z^{s} Z^{s'}
}
where $w_{s, s'}$ is the inverse of the intersection form and $Z^{2k}= -{m_{2k} \over 2}$, $Z^{2k +1} =- i {m_{2k + 1} \over 2} $,
$Z^{n-1} = -{\bar{m} \over 2}$.

$A_{free}$ in the formula above is given by the free energy of the TBA system
\eqn\Afree{\eqalign{
A_{free}^{ff} &= \sum_{s} \int \frac{ d \theta}{2 \pi} |m_{s}| \cosh \theta \log (1 + \tilde{Y}_{s})
+ 2 \int \frac{ d \theta}{2 \pi} |\bar{m}| \cosh \theta \log (1 + \tilde{\bar{Y}}) \cr
m_{s} &= | m_{s} | e^{i \phi_{s}} \quad \bar{m} = |\bar{m}| e^{i \bar{\phi}}\cr
\tilde{Y}_{s}(\theta) &= Y_{s}(\theta + i \phi_{s}), \quad  \tilde{\bar{Y}} = \bar{Y}(\theta + i \bar{\phi})
}}
Notice the appearance of an extra factor of $2$ in front of term containing $\bar{Y}$. This factor
plays an important role in all calculations below.
One sees that this formula contains $2(n-1)$ parameters as it should. The whole
kinematics is encoded in the mass parameters.
We can get rid of the masses and write the area purely in terms of the physical cross ratios which
is described below.

\subsec{Area as the critical value of Yang-Yang functional}

In \refs{\AldayKU} the expression for the area purely in terms of cross ratios
was obtained. The same can be done in the case of form factor  with minor changes
in the  discrete data and the number of Y-functions.

It is convenient to introduce new variables
\eqn\Xwithout{\eqalign{
X_{2k} (\theta) &= Y_{2k} (\theta), \quad X_{2k+1} (\theta) = Y_{2k+1} (\theta - i \frac{\pi}{2}) \cr
X_{n-1} (\theta) &= \bar{Y} (\theta)
}}
for which $X_{s}(\zeta = 1) = \chi_{s}^{+}$ and $X_{s}(\zeta = i) = \chi_{s}^{-}$ where
$\chi_{s}^{\pm}$ are physical cross ratios that define kinematics of the process.

In terms of these variables we can rewrite the Y-system in the form given in equation D.2 of
 \refs{\AldayKU} with the discrete data
$\la s,s+1\ra = -\la s+1,s\ra =(-1)^{s+1}$ and
\eqn\Xdiscreteff{\eqalign{
\Omega(s) &= 1, \quad s=1,..,n-2 \cr
\Omega(n-1) &= 2.
}}
We can use the formulas given in \refs{\AldayKU} for the area
$A_{periods}^{ff} + A_{free}^{ff} = A_{0}^{ff} + YY_{cr}  $ with
\eqn\Xarea{\eqalign{
A_{0}^{ff} &= - \frac{1}{2} \sum_{s,s'=1}^{n-1} w_{s,s'} \log \chi^{+}_{s} \log \chi^{-}_{s'} \cr
YY_{cr}  &= \frac{1}{\pi} \sum_{s = 1}^{n-1} \Omega(s) \int_{l_{s}} \frac{d \theta}{\sinh^2 \theta} {\rm Li}_{2}(- X_{s}(\theta)) +
\cr
 & ~~~~~ + \frac{1}{4 \pi^2 i} \sum_{s,s'=1}^{n-1} \Omega(s) \Omega(s') \la s, s'\ra \int_{l_{s'}} \frac{d \theta'}{\sinh 2 \theta'} \int_{l_{s}} \frac{d \theta}{\sinh 2 \theta}
  \cr &
~~~~~~~~ \frac{1}{\sinh(\theta' - \theta)} \log(1+ X_{s'}(\theta')) \log(1+ X_{s}(\theta)).
}}

The integral equations take the form in equation D.3 of \AldayKU .
To fix the contours of integration we go to the region of parameter space where all cross ratios
 are  very small or very large. In this region the simple relation between masses and cross
 ratios exist so we can fix the contours of integration.
 After it being careful about contribution from the poles in the kernels we
  can continue them to the region of arbitrary cross ratios.

\newsec{Exact solutions }

In this section we consider several simple cases when
the solution of the
Y-system is known and the area can be found exactly. The simplest case is when Y-functions are independent of
the spectral parameter.
These  arise when all masses go to zero. They correspond to the high temperature limit of the TBA.
 Geometrically they correspond to a  polygon with maximal symmetry.
In addition, an exact solution can be found for the form factors case for $n=2$. This solution is very similar to
that of the octagonal Wilson loop \AldayYN .

\subsec{High temperature limit of the form factor Y-system}

Here we consider  solutions of the form factor Y-system which
are independent of the spectral parameter. Geometrically the regular form factor corresponds
to the same zig-zag solution \Zigzag ,  but with a  different choice for the number of cusps
per period.

In the case of amplitudes the answer is known to be
\eqn\RegAmp{\eqalign{
A_{Sinh}^{sa} (n) &= A_{free}^{sa}(n) + (n -2) C_{6} = {\pi \over 4 n} (3 n^2 - 8 n + 4)\cr
A_{free}^{sa}(n) &= \frac{\pi}{6} \frac{(n-3)(n-2)}{n}
}}
where $2n$ is the total number of gluons in the problem.

For the form factor we have
\eqn\Affsihn{\eqalign{
A_{Sinh}^{ff}(n) = A_{free}^{ff}(n) + C_{zz} + (n-1) C_{6}
}}
where $2n$ is again the total number of gluons.
From the definition of the regular form factor we see that
 $A_{Sinh}^{ff}(n) = n C_{zz}$.
Thus, we get the expression for the free energy
\eqn\AcheckAD{\eqalign{
n C_{zz} &= A_{free}^{ff}(n) + C_{zz} + (n-1) C_{6} \cr
A_{free}^{ff}(n) &= (n-1) (C_{zz} - C_{6}) = \frac{\pi}{6} (n-1).
}}

The first check of this formula can be done for the cases when we have only one or two non-zero Y-functions
in the form factor problem. One can notice looking at corresponding  Y-systems \Yamp\ and \Yoperins\ that they
are completely equivalent in these cases in high temperature limit and the following equalities should hold\foot{Using different language it can be understood at the level of Dynkin diagrams as $D_{2}=2 A_{1}$ and $D_{3}=A_{3}$ relations where the form factor Y-system corresponds to $D_{n}$ series and the scattering amplitudes one to $A_{n}$ series \refs{\ZamolodchikovET}.}
\eqn\EqDAfirst{\eqalign{
A_{free}^{ff}(2) &= 2 A_{free}^{sa}(4), \cr
A_{free}^{ff}(3) &= A_{free}^{sa}(6).
}}
Using the formulas \RegAmp\ and \AcheckAD\ one can check that this is indeed true.

More non-trivial check is to reproduce the result for the free energy of regular form factor \AcheckAD\ using the Y-system.
If we take the zig-zag solution and we choose the period to contain $2n$ cusps then the solution of \Yoperins\ is
\eqn\FFreg{\eqalign{
Y_{s} &= s (s+2) \cr
\bar{Y} &= n-1
}}
one can get this solution in two ways: either from geometrical
definition of the Y-functions or by taking large $n$ limit of the regular polygon solution
for $AdS_3$ \YZMb,\YZMc.

The formula for the free energy in this case takes the form \refs{\FendleyXN}:
\eqn\FFregB{\eqalign{
A_{free}^{ff} (n)&= - \frac{1}{2 \pi} \sum_{s=1}^{n-2}(\log(Y_{s}) \log(1+Y_{s})+2 Li_2 (- Y_{s})) \cr
&- \frac{1}{\pi}(\log(\bar{Y}) \log(1+\bar{Y})+2 Li_2 (- \bar{Y}))
}}
inserting values for Y's \FFreg\ one can check that this sum is indeed equal to $\frac{\pi}{6} (n-1)$.

Another check can be done by viewing the form factor as an infinite number of gluons limit of the amplitude
\eqn\LimitConn{\eqalign{
A^{ff}_{Sinh} (n) = \lim_{m \to \infty} \frac{A_{Sinh}^{sa}(n m)}{m} = n C_{zz}
}}
which   also holds.

\subsec{High temperature limit of the ${\rm Z_{m}}$ symmetric Y-system}

Using the known results for the area of regular polygons we can make a   consistency check
for the ${\rm Z_{m}}$ symmetric Y-system.

Namely let us consider a regular polygon with $2 m n$ cusps and view it as a special case of a  ${\rm Z_{m}}$ symmetric one.
By the construction
\eqn\ZMrelations{\eqalign{
m A_{Sinh}^{Z_{m}}(n) &= A_{Sinh}^{sa} (n m) \cr
A_{Sinh}^{sa} (n m) &= A_{free}^{sa}(n m) + (n m -2) C_{6} \cr
A_{Sinh}^{Z_{m}} (n) &= A_{free}^{Z_{m}}(n) + C_{p}(m) + (n-1) C_{6}
}}
where we again consider the limit when all zeroes and pole are far away from each other
to get the second and the third formulas. We introduced the $C_{p}(m)$ for the area of the isolated pole.
It should appear as the area for
the solution of modified Sinh-Gordon with $p(z)={1 \over z}$ and a boundary condition for $\alpha$ given in Appendix B.

By considering \ZMrelations\ for $n=1$ we get
\eqn\ZMpole{\eqalign{
C_{p}(m) = \frac{A_{Sinh}^{sa} (m)}{m} = \frac{A_{free}^{sa}}{m} + \frac{m -2}{m} C_{6} = \frac{\pi (m-2)(3 m - 2)}{4 m^2}.
}}
Putting all together we get
\eqn\ZMfreereg{\eqalign{
A_{free}^{Z_{m}} (n) &= \frac{A_{free}^{sa}(n m)-A_{free}^{sa}(m)}{m} = \frac{\pi}{6} (n-1) (1 - \frac{6}{m^2 n}).
}}

Again it is interesting to reproduce this result directly from the Y-system. In the case of ${\rm Z_{m}}$ symmetric
Y-system \YZM\ the free energy takes the form
\eqn\ZMreghigh{\eqalign{
A_{free}^{Z_{m}}(n) &= - \frac{1}{2 \pi} \sum_{s=1}^{n-2}(\log(Y_{s}) \log(1+Y_{s})+2 Li_2 (- Y_{s})) \cr
&- \frac{1}{2 \pi}(\log(\bar{Y}) \log(1+e^{i \frac{\pi}{m}} \bar{Y})+2 Li_2 (- e^{i \frac{\pi}{m}} \bar{Y})) \cr
&- \frac{1}{2 \pi}(\log(\bar{Y}) \log(1+e^{-i \frac{\pi}{m}} \bar{Y})+2 Li_2 (- e^{- i \frac{\pi}{m}} \bar{Y})).
}}
Plugging the solution \YZMb,\YZMc\ into \ZMreghigh\ we reproduce \ZMfreereg .
Another point is that we expect from the form factor as an infinite number of gluon limit the following equations to be satisfied
\eqn\ZMlimitcheck{\eqalign{
A_{free}^{ff}(n) &= \lim_{m \to \infty} A_{free}^{Z_{m}}(n) \cr
C_{zz} &= \lim_{m \to \infty} C_{p}(m).
}}
One can easily check that this is indeed true.

\subsec{Exact solution for the 4-cusp form factor}

\ifig\Fourcusp{We consider an  operator  going
 into states with 4 gluons which at strong coupling is given by the surface with boundary condition shown in the picture. The zeroth cusp chosen to be at the origin. The momentum of the operator is $q$. The picture corresponds to the case when $\chi^{+} \ll 1$ and $\chi^{-} \gg 1$.}
{\epsfxsize4.0in\epsfbox{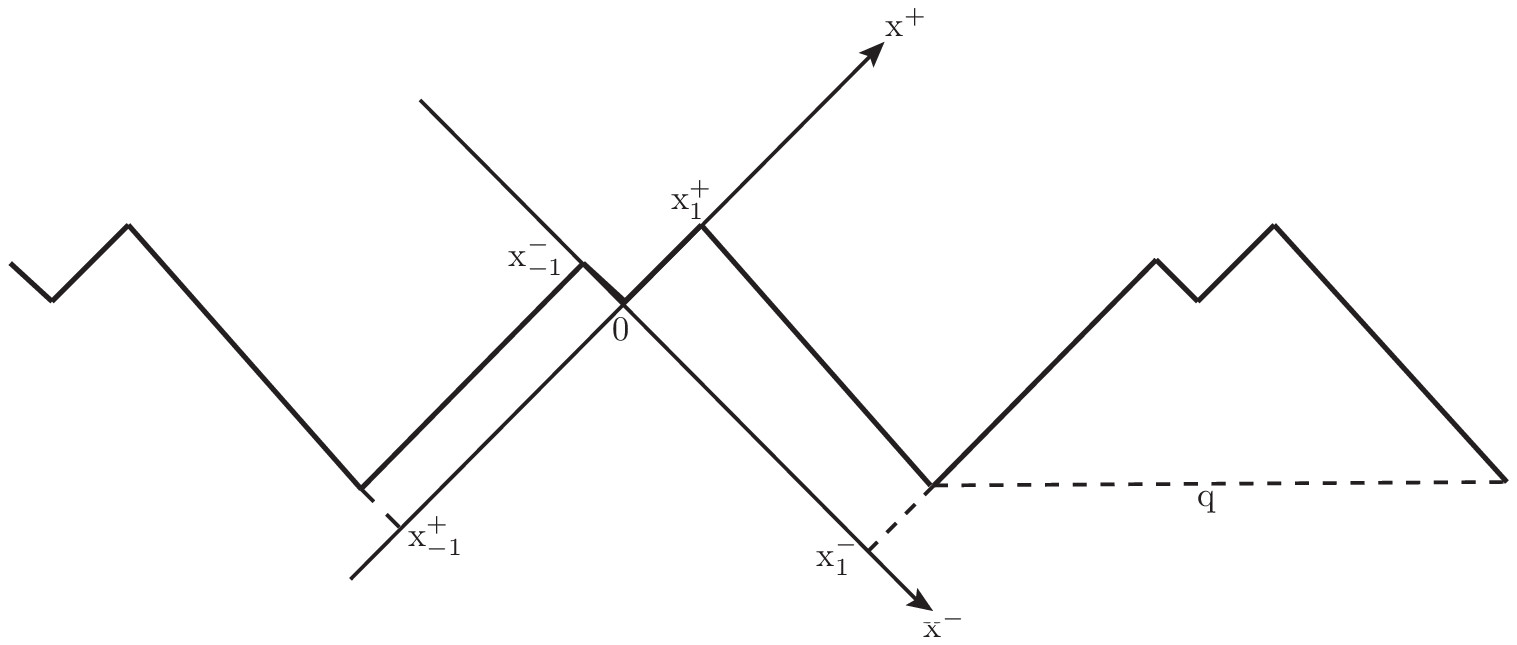}}

Here we give the exact solution for the area in the case of 4-cusp form factor
depicted in \Fourcusp. Here $n=2$ and the  Y-system   is just $\bar{Y}^{+} \bar{Y}^{-} = 1$,  with
the solution $\bar{Y} = e^{Z/\zeta + \bar{Z} \zeta}$. We denote the only cross ratio in the problem
by  $\chi$ with the definition
\eqn\CRfourc{\eqalign{
\bar{Y}(\zeta=1) &= \chi^{+} = {x_{1,0}^{+} \over x_{0,-1}^{+}}, \cr
\bar{Y}(\zeta=i) &=\chi^{-} = {x_{1,0}^{-} \over x_{0,-1}^{-}}.
}}
Due to the fact that $n$ is even in this case we cannot  directly apply formulas we have written above  \ArPieces , \ADivBds .
However, we can derive the correct result by viewing this case as the
 the double soft limit from the $n=3$ form factor problem.
 This is explained in detail in Appendix E.
 The case of 4-cusp form factor is analogous to the octagon up to the several important coefficients and signs.
  Also here we do not subtract from the area $A_{BDS}$ but take the limit directly at the level of $A_{BDS-like}$.

With given definitions the answer for the area is
\eqn\AreaFC{\eqalign{
A &= A_{div} + \tilde{A}_{BDS-like} + R \cr
\tilde{A}_{BDS-like} &= {1 \over 4} ( \log \chi^{-} \log \chi^{+} + \log \chi^{-} \log (1 + \chi^{+})^2 - \log \chi^{+} \log (1 + \chi^{-})^2) \cr
R &= {4 \pi \over 3}+ 2 I \cr
I &= \int_{- \infty}^{\infty} d t \frac{|m| \sinh t}{2 \pi \tanh (2 t + 2 i \phi)} \log (1 + e^{- 2 |m| \cosh t}), \quad \phi \in  (0, {\pi \over 2})
}}
where
\eqn\CRfourcM{
\log \chi^{+} = -|m| \sin \phi, \quad \log \chi^{-} = |m| \cos \phi.
}
\ifig\Iperiodic{Here the complex $t$ plane is presented with the poles located at $- i\phi + i {\pi n \over 2}$. We show contours
$\gamma_{i}$ which enter in the definition of $I_{{\rm periodic}}$.}
{\epsfxsize2.5in\epsfbox{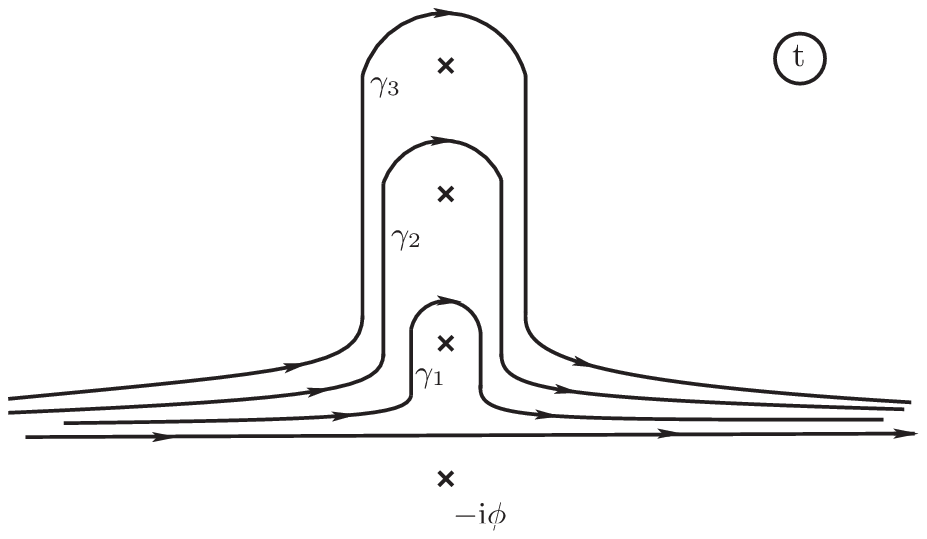}}
Written in this form it is not obvious that the answer is invariant under  cyclic permutations or spacetime parity.
It is due to the fact that as we increase $\phi$ poles can cross the real line which we are integrating over. The deformation of the contour
can be rewritten as the sum of real line integral plus the integral encircling the pole \refs{\AldayYN}.
To make this symmetries manifest  it is useful to define the function
$I_{{\rm periodic}} = {1 \over 4} (I+I_{\gamma_{1}}+I_{\gamma_{2}}+I_{\gamma_{3}})$ where
 $I$ is the integral over the real line and $\gamma_{i}$ are contours shown in \Iperiodic\  with the same integrand
as in \AreaFC. Then $I_{{\rm periodic}}(\phi + {\pi \over 2}) = I_{{\rm periodic}}(\phi)$ and one can show that
\eqn\ResPer{
2 I = 2 I_{{\rm periodic}} -   \tilde{A}_{BDS-like}
}
and the answer for the area can be written as
\eqn\AreaFCb{\eqalign{
A -  A_{div} = {4 \pi \over 3}+ 2 I_{{\rm periodic}}(|m|, \phi).
}}
Written in this form the area exhibits the explicit spacetime parity and cyclicity symmetries.
In the $|m| \to 0$ limit we have $A -  A_{div} = {4 \pi \over 3}+ 2 {\pi \over 12} = {3 \pi \over 2}$ which is the correct
answer for the two copies of zig-zag.

Analogously we can get the critical Yang-Yang functional of the $Z_{m}$ symmetric polygon with $4m$ number of gluons
\eqn\FreeZM{\eqalign{
YY^{Z_{m}}(2) &= \int_{- \infty}^{\infty} d t {|m| \sinh t \over 2 \pi \tanh (2 t + 2 i \phi)}  \log (1 + e^{i {\pi \over m}} e^{- 2 |m| \cosh t})(1 + e^{- i {\pi \over m}} e^{- 2 |m| \cosh t}), \cr
 \phi &\in  (0, {\pi \over 2})
}}
where
\eqn\ZMexact{
\log \chi^{+} = -|m| \sin \phi, \quad \log \chi^{-} = |m| \cos \phi
}
and
\eqn\ZMcross{
\chi^{+} = { \sin {\phi^{+}_{1,0} \over 2} \over \sin {\phi^{+}_{0,-1} \over 2}}, \quad \chi^{-} = { \sin {\phi^{-}_{1,0} \over 2} \over \sin {\phi^{-}_{0,-1} \over 2}}.
}

One can show that at $m=2$ it correctly reproduces the answer for the octagon while in the limit $m \to \infty$
it goes to the solution for the four cusp form factor.

\newsec{Conclusions}

In this paper we have considered the problem of calculating   form factors at strong coupling,
limiting ourselves to $R^{1,1}$ kinematics.

These are given by the area of minimal surfaces in $AdS$ space which
end on a periodic sequence of null segments at the boundary of $AdS$ space. The shape of the sequence is
fixed by the gluon momenta. The operator momentum defines the period of the sequence.

The problem can be reformulated in terms of a flat section problem for the spectral parameter dependent
 flat connection.
The insertion of the operator creates a non-trivial monodromy on the worldsheet which specifies  the behavior
of the connection near the insertion point. This monodromy  characterizes the operator.
 Without an operator insertion (with a trivial monodromy)
 this problem was recently solved using the integrability of classical strings in $AdS$  \refs{\AldayVH}.
  In that case,
 one needs to
solve a set of functional equations for the cross ratios as functions of the spectral parameter with given boundary condition at
$\theta \to \pm \infty$ - the Y-system. Here we extended the previous analysis and derived a set of functional
equations for the insertion of a general operator  \Yoper . In fact, these functional equations should be valid for
general operators, even those dual to semiclassical string states, or the ones described by ``finite gap'' solutions  \refs{\KazakovQF,\DoreyZJ}.
They should even be valid in cases where we have any other structure near $z=0$, such as another Wilson loop, though
in such cases we would probably need more Y-functions to fully specify the system. In other words, the equations
\Yoper\ are a local property of the irregular singularity at infinity and the holonomy around it.

We have then concentrated on    a couple of
 examples when the monodromy does not depend on the spectral parameter.
One corresponds to the case of form factors of operators with conformal dimensions
  small   compared to $\sqrt{\lambda}$. The monodromy can be found explicitly \OpIns . Supplementing
the functional equations \Yoperins\ with boundary conditions using a  WKB analysis, we rewrote them in the form
of integral equations of the TBA form \Yintegral .
The area is then given by the free energy of the TBA system or critical value of Yang-Yang functional.
\Afree ,  \Xarea .

It should be noted that, at leading order, the answer is insensitive to the particular form of the operator.
This has a physical explanation: at strong coupling,  the production of finite number of quanta is equally  strongly suppressed
for all operators \refs{\PolchinskiJW,\HofmanAR}. The dependence on the precise operator, as well as the dependence
on the gluon polarizations should reappear at one loop in the $1/
\sqrt{\lambda}$ expansion.

First the analysis was done for the case when we have an operator and $2 n$ gluons with $n$ being odd. The case
of even $n$'s can be then obtained as a simple double soft limit which we explained both for the case of amplitudes and form factor (see Appendix E).

Using this knowledge we solved exactly the problem for an operator creating
 $4$ gluons. The area is given by \AreaFCb.
Another easily tractable case is the so called high temperature limit. In this case the Y-functions do not depend on the spectral parameter and the solution of the Y-system  can also be found. We used it as a non-trivial check of our formulas.

This analysis can be extended in a few other directions.
The most obvious one is to generalize it  to the case of $AdS_{5}$ or full $R^{1,3}$ kinematics.
Another direction is to consider the insertion of more general operators which are given by
classical string solutions at strong coupling
and for which the anomalous dimension at leading order is non-zero.
One famous example of such a solution is GKP  string \refs{\GKP}. In this case we will have the
functional equations derived in \Yoper\ where ${\rm Tr}[\tilde \Omega]$ should be the one corresponding
to the operator. What remains to be done is the derivation of the expression for the area. This
could require the introduction of extra functions.
Another direction is to extend the present analysis to the
 case of an insertion of several operators which can then be related
 to the problem of the calculation of   correlation functions
 in ${\cal N}=4$ SYM at strong coupling. One more question is whether
 there is  a  form factor/Wilson line analogue of the scattering amplitudes/Wilson loop duality at weak coupling.

\newsec{Acknowledgments}

We thank Benjamin Basso for collaboration at the beginning of the project. We also
thank Luis F. Alday, Davide Gaiotto, Pedro
Vieira and Amit Sever,  for very useful
comments and suggestions, as well as for collaboration in closely related issues.

\appendix{A}{Derivation of the monodromy}

Here we present the derivation for the monodromy in the case of an operator insertion. We use
the explicit solution of the section problem in the vicinity of operator insertion. This
is given by the following equations
\eqn\Asect{\eqalign{
\partial_{z} \psi + B_{z}(\zeta) \psi &= 0 \cr
\partial_{\bar{z}} \psi + B_{\bar{z}}(\zeta) \psi &= 0
}}
where
\eqn\Aconn{\eqalign{
B_{z} = \left( \matrix{ \frac{1}{2} \partial_{z} \alpha & -\frac{1}{\zeta} e^{\alpha} \cr -\frac{1}{\zeta} e^{-\alpha} p(z) & -\frac{1}{2} \partial_{z} \alpha} \right), \quad B_{\bar{z}} = \left( \matrix{ -\frac{1}{2} \partial_{\bar{z}} \alpha & - \zeta e^{-\alpha}  \bar{p}(\bar{z}) \cr - \zeta e^{\alpha} & \frac{1}{2} \partial_{\bar{z}} \alpha}.
\right)}}
are components of the flat connection which appears in the reduced formalism. The connection is defined in terms
of a solution of modified Sinh-Gordon $\alpha(z, \bar{z})$ and the polynomial $p(z)$ that controls the kinematics of the process.
It is convenient to make a gauge transformation
\eqn\Agauge{\eqalign{
\partial_{z} + \hat{B}_{z} &= g[\partial_{z} + B_{z}]g^{-1}\cr
\partial_{\bar{z}} + \hat{B}_{\bar{z}} &= g[\partial_{\bar{z}} + B_{\bar{z}}]g^{-1} \cr
\hat{\psi} &= g \psi
}}
with
\eqn\AgaugeExp{\eqalign{
g = \left( \matrix{ (\frac{z}{\bar{z}})^{-1/4} & 0 \cr  0 & (\frac{z}{\bar{z}})^{1/4} }\right).
}}
In the formulas for the connection \Aconn\ we substitute
\eqn\AmodSG{\eqalign{
\alpha (z, \bar{z}) &= - \frac{1}{2} \log(z \bar{z} \log^2(z \bar{z})) \cr
p(z) &= \frac{a_{-1}}{z} + a_{0} + a_{1} z + ... \cr
\bar{p}(\bar{z}) &= \frac{a_{-1}}{\bar{z}} + a_{0} + a_{1} \bar{z} + ...
}}
and solve the section problem in the vicinity of the origin $z= \rho e^{i \phi}$, $\rho \to 0$.

At the leading order in $\rho$ we get the following orthonormal pair of solutions
\eqn\Apair{\eqalign{
\hat{\psi}_{1} &= \frac{c}{\sqrt{\log(z \bar{z})}} \left( \matrix{ 1 \cr \zeta  } \right) \cr
\hat{\psi}_{2} &= \frac{1}{c \sqrt{\log(z \bar{z})}} \left( \matrix{ - \frac{\log z}{\zeta} \cr \log \bar{z}} \right).
}}
We define the monodromy as
\eqn\Amono{\eqalign{
\hat{\psi}_{a} (z e^{2 \pi i}) = \Omega_{a}^{~ b} \hat{\psi}_{b} (z).
}}
and see that
\eqn\AmonoB{\eqalign{ 
\Omega = \left( \matrix{ 1 & 0 \cr  -\frac{2 \pi i}{\zeta c^2} & 1 } \right).
}}
Note that the monodromy is independent of the polynomial, as expected.
 This monodromy, found near the origin,  is  the same on the whole complex plane.
By choosing the normalization constant to be
\eqn\Anorm{\eqalign{
c^{2}(\zeta) = - \frac{2 \pi i}{\zeta q}.
}}
we get the monodromy that we used in the main body of the text
\eqn\Aend{\eqalign{
\Omega(\zeta) = \left( \matrix{1 & 0 \cr q & 1 } \right).
}}

\appendix{B}{${\rm Z_{m}}$ symmetric polygons}

Let's consider the scattering amplitudes problem in the case of
most general ${\rm Z_{m}}$ symmetric polygon. In terms of modified Sinh-Gordon
it is given by the following problem
\eqn\BmapFrom{\eqalign{
\partial_{z} \partial_{\bar{z}} \alpha(z, \bar{z}) &- e^{2 \alpha(z,\bar{z})}+|p(z)|^2 e^{- 2 \alpha(z, \bar{z})} = 0 \cr
p(z) &= m^2 z^{m-2} ( 1 + a_{0} z^m + ... + a_{n-2} z^{m (n-1)} )\cr
\alpha & \quad {\rm regular} , \quad z \to 0 \cr
\hat{\alpha} &\to 0, \quad |w| \to \infty.
}}

Now we can make conformal transformation $z = \tilde{z}^{1/m}$ which allows us to focus on the one period.
We get
\eqn\BmapTo{\eqalign{
\partial_{z} \partial_{\bar{z}} \alpha(z, \bar{z}) &- e^{2 \alpha(z,\bar{z})}+|p(z)|^2 e^{- 2 \alpha(z, \bar{z})} = 0 \cr
p(z) &= \frac{1}{z} + a_{0} + ... +a_{n-3} z^{n-3} + a_{n-2} z^{n-2} \cr
\alpha &= l \log(z \bar{z}) + {\rm regular} , \quad z \to 0 \cr
\hat{\alpha} &\to 0, \quad |w| \to \infty \cr
l &= \frac{1}{2 m} - \frac{1}{2}
}}
here $l = \frac{1}{2 m}- \frac{1}{2} $. This makes a connection with the problem
considered recently by Lukyanov and Zamolodchikov \LukyanovRN   . In particular they argue that
for this kind of problem there exist smooth limit $l \to - \frac{1}{2}$ which is
\eqn\BmapLimit{\eqalign{
\partial_{z} \partial_{\bar{z}} \alpha(z, \bar{z}) &- e^{2 \alpha(z,\bar{z})}+|p(z)|^2 e^{- 2 \alpha(z, \bar{z})} = 0 \cr
p(z) &= \frac{1}{z} + a_{0} + ... +a_{n-3} z^{n-3} + a_{n-2} z^{n-2} \cr
\alpha &= - \frac{1}{2} \log(z \bar{z} \log^2(z \bar{z})) + {\rm less ~singular}, \quad z \to 0 \cr
\hat{\alpha} &\to 0, \quad |w| \to \infty
}}
and corresponds to operator insertion problem considered above.

\appendix{C}{Computation of $A_{cut-off}$ for $n$ odd}

We are interested in the calculation of the integral
\eqn\Carea{\eqalign{
A_{cut-off} = 4 \int_{\Sigma_0, z_{AdS}> \epsilon}d^2 w.
}}

The algorithm was first developed in \refs{\AldayYN}, to which we refer the reader for notation and further discussion.
Here we just   repeat the analysis keeping in mind that we have
non-trivial monodromy and check that we get essentially the same as in \AldayYN .

Below we think about any flat section as $2 \times 2$ matrix with unit determinant $\psi_{\alpha a}$
where $\alpha$ is inner $SL(2,R)$ and $a$ target $SL(2,R)$ indices.

If we consider $z$ plane with $\psi_{\zeta}(z)$ being exact solution of the problem
then the target space solution is
\eqn\Cy{\eqalign{
Z &= \psi^{T}_{\zeta=1} U \psi_{\zeta = i}.
}}
As we go around the $z$-plane $z \to e^{2 \pi i} z$ the solution transforms as
\eqn\Ctrans{\eqalign{
Z &\to \hat{\Omega} Z \hat{\Omega}^{T} \cr
\hat{\Omega} &= \left( \matrix{ 1 & 0 \cr q & 1 } \right)
}}
this monodromy corresponds to the translation of the Poincare coordinate $x \to x + q$.
Recall that having $q$ in lower left position is important. Otherwise
the monodromy does not correspond to $x \to x + q$. This will be important below.

Thus, we have
\eqn\Csect{\eqalign{
\psi(z e^{2 \pi i}) = \psi(z) \hat{\Omega}^{T}
}}

As in \AldayYN , we
go to the $w$-plane with the gauge transformation being not single-valued in
the $z$-plane
\eqn\Cgauge{\eqalign{
\hat{\psi} (w) = g(z) \psi(z).
}}
As $z \to e^{2 \pi i} z $, $ w \to e^{i \pi n} w$ and we have
\eqn\Ckey{\eqalign{ 
\hat{\psi} (w e^{i \pi n}) = e^{i \frac{\pi}{2} (n-2) \sigma_2} \hat{\psi}(w) \hat{\Omega}^{T}
}}
for the details see \refs{\AldayYN} the only difference is the appearance of the $\hat{\Omega}^{T}$ which
was the identity matrix for the case of amplitudes.

The exact solution is smooth. However,
as we go to the large $w$ ($\hat{\alpha} \to 0$) region the exact solution is approximately equal (in the sense of asymptotic expansion) to (for the left problem)
\eqn\Cappr{\eqalign{
\hat{\psi}_{appr}(w) = \left( \matrix{ b_{1 1} e^{w + \bar{w}} & b_{1 2} e^{w + \bar{w}} \cr b_{2 1} e^{ - w - \bar{w}}& b_{2 2} e^{ - w - \bar{w}}}\right)
}}
where coefficients $b_{i j}$ are different in each Stokes sector. For the given Stokes sector we denote them as $b_{i j}^{k}$ where $k$ is the number
of the Stokes sector.

If we analytically continue $\hat{\psi}_{appr}(w)$ as soon as we cross a Stokes line it is not the right form of the asymptotic form of the
exact solution. Every time we cross a Stokes line the coefficients of asymptotic expansion jump and this information is encoded in Stokes matrices.

Let's consider
\eqn\Cafter{\eqalign{
\hat{\psi}_{appr}^{n+1}(w) = \left( \matrix{ b_{1 1}^{n+1} e^{w + \bar{w}} & b_{1 2}^{n+1} e^{w + \bar{w}} \cr b_{2 1}^{n+1} e^{ - w - \bar{w}}& b_{2 2}^{n+1} e^{ - w - \bar{w}}}
\right)}}
which corresponds to the correct asymptotic form of the solution in $(n+1)$-th Stokes sector.

Now let's take the large $w$ limit in the \Ckey. We get for odd $n$
\eqn\Cimport{\eqalign{ 
\hat{\psi}_{appr}^{n+1}(-w) = e^{i \frac{\pi}{2} (n-2) \sigma_2} \hat{\psi}_{appr}^{1}(w) \hat{\Omega}^{T}.
}}
From this equation one can get the relation between $b_{i j}^{n+1}$ and $b_{i j}^{1}$.
For the right problem all formulae are completely the same.

The central objects for the computation are $\delta u_{i}$
and $\delta v_{i}$ in \AldayYN.
In our notations they are defined as follows
\eqn\Csmallpieces{\eqalign{
\delta u_{2k + 1} &= - \log ( b^{2k+1}_{11} \tilde{b}^{2k+1}_{11}),\quad \delta u_{2k} = - \log ( b^{2k}_{21} \tilde{b}^{2k}_{21}),\cr
\delta v_{2k + 1} &= - \log ( - b^{2k+1}_{21} \tilde{b}^{2k+1}_{11}), \quad \delta v_{2k} = - \log ( b^{2k}_{11} \tilde{b}^{2k}_{21}).
}}

From the \Cimport\ both for amplitudes and form factor one gets
\eqn\CbdsOne{\eqalign{ 
\delta u_{n+1} &= \delta u_{1} \cr
\delta v_{n+1} &= \delta v_{1}.
}}

Equations \CbdsOne\ are the same as in the case of amplitudes.
Thus, we have exactly the same formula for the area as in the case of amplitudes in terms of $(\delta u_{i},\delta v_{i})$.

After noticing the fact that the formula contains only coordinate difference between two consecutive cusps
we conclude that the formula for $A_{cut-off}$ part in terms of $l^{+}_{i}$ and $l_{j}^{-}$ for the form factor is completely identical to the amplitudes' one. Except that instead of $x^{\pm}_{n} = x^{\pm}_{0}$ we have $x^{\pm}_{n}=x^{\pm}_{0}+q$ which does not affect the fact that $l^{\pm}_{i+n} = l^{\pm}_{i}$.
The final formula is given by $A_{BDS-like}$ in equation  (5.8) in \AldayYN .

\appendix{D}{Integral equations for complex masses}

Starting from the case of real masses we can introduce phases which modifies
the form of integral equations.

For $m_{s} = | m_{s} | e^{i \phi_{s}}, \quad \bar{m} = |\bar{m}| e^{i \bar{\phi}}$ we have
\eqn\Dkernel{\eqalign{
K_{s,s'}(\theta) &= \frac{1}{2 \pi \cosh (\theta - \theta' + i \phi_{s} - i \phi_{s'})} \cr
\tilde{Y}_{s}(\theta) &= Y_{s}(\theta + i \phi_{s}), \quad  \tilde{\bar{Y}} = \bar{Y}(\theta + i \bar{\phi})
}}
and for $|\phi_{s} - \phi_{s+1}| < \frac{\pi}{2}$ we have
\eqn\Dint{\eqalign{ 
\log \tilde{Y}_{s} &= - |m_{s}| \cosh \theta + K_{s,s+1} \star \log (1 + \tilde{Y}_{s+1})\cr
 &+ K_{s,s-1} \star \log (1 + \tilde{Y}_{s-1}), \quad s=1,..., n - 3 \cr
\log \tilde{Y}_{n-2} &= - |m_{n-2}| \cosh \theta + K_{n-2,n-3} \star \log (1 + \tilde{Y}_{n-3}) \cr
 &+ 2 K_{n-2,\bar{\phi}} \star \log (1 + \tilde{\bar{Y}})  \cr
\log \tilde{\bar{Y}} &= - |\bar{m}| \cosh \theta + K_{n-2,\bar{\phi}} \star \log (1 + \tilde{Y}_{n-2})
}}
as soon as we cross the lines $|\phi_{s} - \phi_{s+1}| = \frac{\pi}{2}, \frac{3 \pi}{2},...$ we get the contribution
from the pole in the kernel and the form of integral equation will change. This was discussed in detail in \refs{\AldayVH}.

\appendix{E}{Calculation of $N=4k$ gluon amplitudes by taking the double soft limit}

Let us consider the amplitude for $2 n$ gluons with $n$ being odd and then
take the double soft limit to get $2(n-1)$ gluons amplitude. This step is necessary since we have
derived the full expression for the area only for $n$ odd. This is a quick way to get the result
when $n$ is even.

Here we present formulas for the area which depend only on the cross ratios.
If we introduce the new variables
\eqn\Xdef{\eqalign{
X_{2k} (\theta) &= Y_{2k} (\theta) \cr
X_{2k+1} (\theta) &= Y_{2k+1} (\theta - i \frac{\pi}{2}) = Y^-_{2k+1}\cr
X_{n-1} (\theta) &= \bar{Y} (\theta)
}}
We choose   integration contours  $l_{\gamma}$  in a canonical way so that all
$Z_{\gamma}/e^{\theta'}$ are real and negative
\eqn\Contour{\eqalign{
l_{2k} &= \left( i \phi_{2k} - \infty, i \phi_{2k} + \infty \right) \cr
l_{2k+1} &= \left( i (\phi_{2k+1}+\frac{\pi}{2}) - \infty, i (\phi_{2k+1}+\frac{\pi}{2}) + \infty \right) \cr
l_{n-1} &= \left( i \bar{\phi} - \infty, i \bar{\phi} + \infty \right)
}}
we get that the Y-system can be rewritten as follows
\eqn\XYsystem{\eqalign{
\log X_{s}(\theta) = Z_{s} e^{- \theta} + \bar{Z}_{s} e^{\theta} + \frac{1}{2 \pi i} \sum_{r=1}^{n-1} \Omega(r) \theta^{s r} \int_{l_{r}} \frac{d \theta'}{\sinh(\theta' - \theta)} \log(1 + X_{r}(\theta'))
}}
where $\theta^{s r}$ \foot{$\theta^{s r} = (-1)^{s+1} (\delta_{s+1,r} + \delta_{s-1,r})$} is the intersection form. Or, using the notation  of \AldayKU ,
\eqn\XYsystemB{\eqalign{
\log X_{s}(\theta) = Z_{s} e^{- \theta} + \bar{Z}_{s} e^{\theta} + \frac{1}{2 \pi i} \sum_{s'=1}^{n-1} \Omega(s') \la s, s'\ra \int_{l_{s'}} \frac{d \theta'}{\sinh(\theta' - \theta)} \log(1 + X_{s'}(\theta'))
}}
with $\la s,s+1\ra = -\la s+1,s\ra =(-1)^{s+1}$. Here $Z_{2k}= -{m_{2k} \over 2}$, $Z_{2k +1} =- i {m_{2k + 1} \over 2} $.

The case of amplitudes then corresponds to \foot{ This $\Omega$ should not be confused with the
monodromy $\Omega$.}
\eqn\XdiscreteSA{\eqalign{
\Omega(s) &= 1, \quad s=1,..,n-3 \cr
}}
and the case of the form factor to
\eqn\Xdiscreteff{\eqalign{
\Omega(s) &= 1, \quad s=1,..,n-2 \cr
\Omega(n-1) &= 2.
}}

The full answer for the area then takes the form
\eqn\AreaFull{\eqalign{
A = A_{div} + A_{BDS-like} + A_{0} + YY_{cr} + C_{0}
}}
here $C^{0}$ is the constant which comes from the difference between $A_{Sinh}$ and $A_{free}$ and is given
\eqn\AreaConstant{\eqalign{
{\rm Scattering ~Amplitudes: } ~~~~C_{0}^{sa} &= {7 \pi \over 12} (n-2) \cr
{\rm Form ~Factor: }~~~~ C_{0}^{ff} &= {3 \pi \over 4}+{ 7\pi \over 12} (n-1).
}}
$YY_{cr}$ is the critical value of the Yang-Yang functional that depends only on Y-functions.
$A_{div}$ and $A_{BDS-like}$ are known and can be found in \refs{\AldayYN} (equation 5.8).

Remember that the phases of the mass parameters entered in the equations for the area as a choice of contours
of integration. To get rid of masses completely we would like to go to the region where all masses parameters are big $m_{s} >> 1$
and where there is a simple relation between the masses and the physical cross ratios. We
define $\hat m_s$  via $Z_s = - \hat m_s/2$ ( $\hat m_{2k} = m_{2k}$, $\hat m_{2k+1} = i m_{2 k +1} $).
We write $\hat m_s = |\hat m_s| e^{ i \hat \phi_s} $
\eqn\CRmass{\eqalign{
\log \chi_s^+ = &  - |\hat m_s| \cos \hat \phi_s ~,~~~~~~~~~~ \log \chi_s^- = - |\hat m_s| \sin \hat
\phi_s
}}
We choose $0 < \hat \phi_{2k+1} < {\pi \over 2 }$ and $- {\pi \over 2 } < \hat \phi_{2 k} < 0 $.
With this choice we do not cross any poles when we express masses through physical cross ratios.

Now we want to take the double soft limit. We fix all phases and send $|\hat{m}_{1}| \to \infty$.
This corresponds to $x_{-1}^{\pm} - x_{0}^{\pm} = \epsilon^{\pm} \to 0$. In this case $\chi_{1}^{\pm} \to 0$ equivalently $X_{1}(\zeta) \to 0$.

When we take the double soft limit,  the amplitude $A - A_{div}$ behaves in a known way
\eqn\SoftL{
A^{2 n}(...,x^{\pm}_{-1},x^{\pm}_{0},...) \to {1 \over 4} \log{{\eps^{+}\over x^{+}_{01}}} \log{{\eps^{-}\over x^{-}_{-2 0}}}+ {7 \pi \over 12} + A^{2 n -2}(...,x^{\pm}_{0},...).
}
we write explicitly ${7 \pi \over 12}$ due to the fact that as one of the masses goes to infinity we get the contribution from the isolated zero to the finite part of the answer which is not related to the amplitude we are interested in.

We rewrite the amplitude as
\eqn\Remaind{
A = A_{div} + A_{BDS} + R
}
where
\eqn\RemaindB{
R = A_{BDS-like} - A_{BDS} + A_{0} + YY_{cr} + C_{0}
}
Then the behavior of \SoftL\ is totally captured by $A_{BDS}$ while $R$ being the function of only cross ratios
has a smooth limit which we will study. Under the soft limit $YY_{cr}$ changes trivially simply by setting
$Y_{1}$ to zero in the formulas. The same happens with $C_{0}(n) \to C_{0}(n-1)$ in \AreaFull .

However for the part $(A_{BDS-like} - A_{BDS}) + A_{0}$ more interesting things happen. Namely each of two pieces
has the part that diverges in the limit but as all divergencies are already captured by $A_{BDS}$ they should cancel.

We will see that the structure of each term is such that allows one to take the limit easily.

\subsec{Soft limit of the $A_{0}$}

Here we discuss the behavior of the $A_{0}$ when we take the soft limit.
We have that
\eqn\Azero{\eqalign{
A_{0}^{n} &= - \frac{1}{2} \sum_{s,s'=1}^{n-3} w_{s,s'}^{n} \log \chi^{+}_{s} \log \chi^{-}_{s'} \cr
&= - \frac{1}{2} \sum_{s'=1}^{n-3} w_{1,s'}^{n}( \log \chi^{+}_{1} \log \chi^{-}_{s'} - \log \chi^{+}_{s'} \log \chi^{-}_{1}) + \tilde{A}_{0}^{n-2}
}}
\eqn\AzeroEven{\eqalign{
\tilde{A}_{0}^{n-1} = \tilde A_0^{n-2} =  - \frac{1}{2} \sum_{s,s'=1}^{n-5} w_{s,s'}^{n-2} \log \chi^{+}_{s+2} \log \chi^{-}_{s'+2}
}}
where we used that $w_{s,s'} = - w_{s',s}$, $w_{2,s'>1}=0$ and that $w^{n}_{s+2,s'+2} = w^{n-2}_{s,s'}$ where $w^{n-2}_{s,s'}$ is the inverse intersection form for $2(n-2)$ gluons problem which exists due to the fact that $(n-2)$ is again odd.

Thus, we see that after taking the soft limit $A_{0}^{n} \to \tilde{A}_{0}^{n-1}$ as first part cancels the same one from $A_{BDS-like} - A_{BDS}$.

\subsec{Soft limit of the $A_{BDS-like} - A_{BDS}$}

\ifig\SoftBDS{In (a) we arrange the $x^{\pm}$ coordinates of the original polygon with $2 n$ cusps
and $n$ being odd. Then we take the double soft limit by sending
 $x^{\pm}_{n} \to x_{1}^{\pm}$. Here we draw only the $x^+_i$ positions, for $x_i^-$ we have the same.
 The resulting polygon with $2(n-1)$ number of cusps is shown in (b).
 In (c) we show the auxiliary Wilson polygon with $2(n-2)$ cusps which we
 use to evaluate $c_{i j}^{aux}$. They are uniquely defined because $(n-2)$ is odd.
 In (d) we illustrate the definition of $c_{j 1}^{left}$ (in red) and of
 $c_{j 1}^{right}$ (in blue). In (e) and (f) we illustrate the definition of
 $d_{j 1}^{left}$ (in red) and of $d_{j 1}^{right}$ (in blue). For (d-f) the cross ratio
 is formed by starting with $x_{j 1}^{\pm}$ in the numerator and then following the depicted path,
 putting the next difference in the denominator, the following in the numerator, etc.
\ } {\epsfxsize3.0in\epsfbox{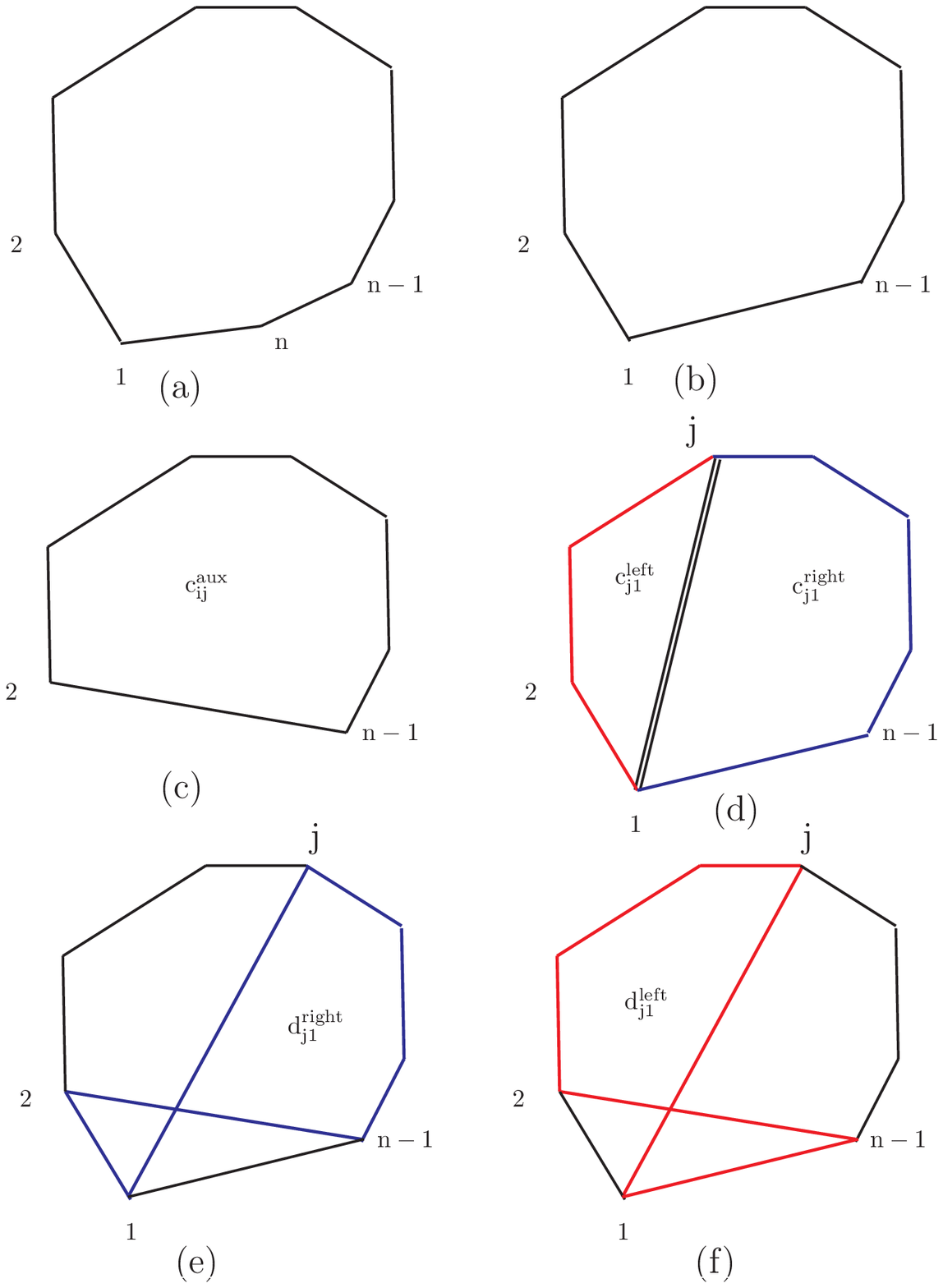}}

As was shown in \refs{\AldayYN}
\eqn\Res{
\Delta A_{BDS} = A_{BDS-like} - A_{BDS} =
{1 \over 4} \sum_{i=1}^{n}\sum_{j=1, j\neq i, i-1}^{n} \log {c_{i,j}^{+} \over c_{j+1,i}^{+}} \log {c_{i-1,j}^{-} \over c_{i,j}^{-}}
}
where $c_{i,j} = c_{j,i}$ are the uniquely defined cross ratios. They start with the factor $x_{ij}$ in the
numerator and are then completed to cross ratios by neighboring differences such as $x_{i,i+1}$.
 We also
define $c_{i,j}=1$ for $|i-j| \leq 1 $.  For $n$ odd this defines them uniquely. Not all these
cross ratios are functionally independent.

When we take the soft limit each $c_{i j}$ can be finite or go to zero or infinity.
Each $c_{ij}$ can be rewritten   as $c_{i, j} = \chi_{1}^{q_{i,j}} \times \hat{c}_{ij}$ where
$q_{i,j}$ is an integer equal to $-1,0,1$ depending on the particular cross ratio. Here $\chi_1$ is the
cross ratio appearing in our basis of Y-functions that we are sending to zero.
 $\hat{c}_{i,j}$ are cross ratios which we describe below.
Thanks to the logs each term in the sum then can be rewritten as
\eqn\FinPart{\eqalign{
\log c_{a_1,b_1}^{+} \log c_{a_2,b_2}^{-} =& \log (\chi_{1}^{+})^{q_{a_1,b_1}}
\log (\chi_{1}^{-})^{q_{a_2,b_2}} + \log (\chi_{1}^{+})^{q_{a_1,b_1}} \log \hat{c}_{a_2,b_2}^{-}\cr
& + \log \hat{c}_{a_1,b_1}^{+} \log (\chi_{1}^{+})^{q_{a_2,b_2}} + \log \hat{c}_{a_1,b_1}^{+} \log \hat{c}_{a_2,b_2}^{-}.
}}
We see that each term in the sum produces a  unique finite piece. Divergencies at the same time should cancel among each other and with the terms
coming from the $A_{0}$. Thus, under the soft limit
 $\Delta A_{BDS}^{n} \to \Delta \tilde{A}_{BDS}^{n-1} $ where
\eqn\ResBDStil{
A^{\tilde n-1}_{BDS} - A^{  n -1}_{BDS-like} = \Delta \tilde{A}_{BDS}^{n-1} = {1 \over 4} \sum_{i=1}^{n}\sum_{j=1, j\neq i, i-1}^{n} \log {\hat{c}_{i,j}^{+} \over \hat{c}_{j+1,i}^{+}} \log {\hat{c}_{i-1,j}^{-} \over \hat{c}_{i,j}^{-}}
}
and $\hat{c}_{i-1,j}$ will be defined below without any reference to the original polygon.
This is the definition of $A^{n-1}_{BDS-like}$ when $n-1$ is even.

To explain pictorially how they are defined it is useful to change the counting of cusps from $-{n-1 \over 2},...,0,...,{n-1 \over 2}$ which is convenient when we think in terms of Y-system to $1,..,n$ with the first cusp corresponding to zeroth one. Then taking the double soft limit corresponds to $x_{n} \to x_{1}$ and we end up with the polygon labeled by $(x_{1},...,x_{n-1})$.

The definition of the $\hat{c}_{i j}$ is simple in terms of the auxiliary odd cusp polygon which is formed by $(x_{2},...,x_{n-1})$ points. Remember the point $x_{1}$ was special in taking the soft limit. Then $\hat{c}_{i,j}$ with $i,j \neq 1,n$ are just $c_{i j}^{aux}$ cross ratios which we defined before for the general odd $n$ polygon applied to the auxiliary polygon. See \SoftBDS\ (c).

Next $\hat{c}_{j,1}$ and $\hat{c}_{n,j}$  with $j$ being even
are given be $c_{j,1}^{left}$ and $c_{j,1}^{right}$ where these
are cross ratios defined as in the case of odd polygon but applied for the even polygon
that  we get after taking the soft limit. For the even polygon there are two cross ratios we can
make from a given distance (we can close it on the left or on the right,
left or right are defined relative to the vector going from $i$ to $j$ where $i<j$). See \SoftBDS\ (d).
Notice that $\hat{c}_{n-1,1} = c_{n-1,1}^{left} = {1 \over c_{2,1}^{right} } = {1 \over \hat{c}_{n,2} }$ being equal to ``round'' cross ratio that is specific only for $n$ even cases and contain only consecutive cusps.

The last parts are $\hat{c}_{j,1}$ and $\hat{c}_{n,j}$  with $j$ being odd. These are given by $d_{j,1}^{right}$ and $d_{j,1}^{left}$. For the definition see \SoftBDS\ (e-f). Their new feature is flip between non-consecutive cusps before closing the loop.

Putting all together we have
\eqn\SumChat{\eqalign{
\hat{c}_{i j} = \left( \matrix{1 & . & . & . & . & ... \cr 1 & 1 & . & . & . & ...\cr d_{3,1}^{right} & 1 & 1 & . & . & ...\cr c_{4,1}^{left} & c_{4,2}^{aux} & 1 & 1 & . & ... \cr . & . & . & . & . & ... \cr c_{2k,1}^{left} & c_{2k,2}^{aux} & c_{2k,3}^{aux} & c_{2k,4}^{aux} & c_{2k,5}^{aux} & ... \cr d_{2k+1,1}^{right} & c_{2k+1,2}^{aux} & c_{2k+1,3}^{aux} & c_{2k+1,4}^{aux} & c_{2k+1,5}^{aux} & ...\cr . & . & . & . & . & ... \cr  1 & c_{2,1}^{right} & d_{3,1}^{left} & c_{4,1}^{right} & d_{5,1}^{right} & ...  } \right) \matrix{{\rm 1} \cr {\rm 2} \cr {\rm 3} \cr {\rm 4} \cr ... \cr 2k \cr 2k+1 \cr ... \cr {\rm n}}
}}

\vskip3em
\vbox{
\begintable
{\rm Case of} $2(n-1)$ {\rm gluons}  | $\hat{c}_{i,j},~~~i>j$
  \cr
  $i,j \neq 1,n.$   | $c^{aux}_{i,j}$
 \cr
  $j = 1;~~~i= 2 k.$   | $c^{left}_{i,1} = { x_{i,1} x_{i-1,i-2} ... x_{3, 2}\over x_{i,i-1} ... x_{2, 1}}$
 \cr
  $j = 1;~~~i= 2 k + 1.$   | $d^{right}_{i,1} = { x_{i,1} x_{i+2,i+1} ... x_{n-1,2}\over x_{i+1,i} ... x_{2, 1}}$
  \cr
  $i = n;~~~j= 2 k.$   | $c^{right}_{j,1} = { x_{j,1} x_{j+2,j+1} ... x_{n-1, n-2}\over x_{j+1,i} ... x_{n-1, 1}}$
 \cr
  $i = n;~~~j= 2 k + 1.$   | $d^{left}_{j,1} = { x_{j,1} x_{j-1,j-2} ... x_{n-1, 2}\over x_{j,j-1} ... x_{n-1, 1}}$
  \endtable}
\centerline{{ Table~2:\/} Explicit expressions for the $\hat{c}_{i j}$ cross ratios when $n$ is even.}
\centerline{These are represented graphically in figure \SoftBDS\ and should be inserted in \ResBDStil.}
\medskip

To summarize let's give once more the expression for the full area for the given even $\tilde{n} = n-1$
\eqn\AreaFulleven{\eqalign{
A = A_{div} + A_{BDS} + \Delta \tilde{A}_{BDS}^{\tilde{n}} + \tilde{A}_{0}^{\tilde{n}} + YY_{cr} + C_{0}. \cr
}}
with $\Delta \tilde{A}_{BDS}^{\tilde{n}}$ given by \ResBDStil\
 and $\tilde{A}_{0}^{\tilde{n}}$ by \AzeroEven\ where we have chosen the first
cusp as being special. One can symmetrize the expression over all cusps to consider all cusps on an equal footing.

In the case of form factors there is slight difference in taking the soft limit which is due to the fact that anomalous conformal Ward identity \refs{\SokWard} acts on the infinite number of points of the periodic polygon and we have no analogue of $A_{BDS}$ which solves it and is defined over one period. In that case we take the soft limit directly at the level of $A_{BDS-like}$ all the rest is completely the same.

\subsec{The octagon}

Let's apply formulas given above to get the answer for the octagon $n=4$ as
a soft limit from the decagon $n=5$. A lot of simplifications occur in this
first non-trivial case for the amplitudes.

First, notice that $\tilde{A}_{0}$ is equal to zero in this case.

As a next step let's analyze the $\Delta \tilde{A}_{BDS}$. Because of few number of cusps there
is no possibility to form a cross ratios using auxiliary Wilson polygon. Thus, the $\hat{c}_{i j}$
take the following form
\eqn\Cij{
\hat{c}_{i j} = \left( \matrix{1 & . & . & . & . \cr 1 & 1 & . & . & . \cr d_{3,1}^{right} & 1 & 1 & . & . \cr c_{4,1}^{left} & 1 & 1 & 1 & . \cr 1 & c_{2,1}^{right} & d_{3,1}^{left} & 1 & 1 } \right)
}
In terms of the cross ratio $\chi$ defined in \refs{\AldayYN} these are
\eqn\Chat{\eqalign{
d_{3,1}^{right} &= 1 + \chi, \quad d_{3,1}^{left} = {1 + \chi \over \chi } \cr
c_{4,1}^{left} &= {1 \over c_{2,1}^{right}} = \chi
}}
plugging it to the formula \ResBDStil\ one gets that $\Delta \tilde{A}_{BDS} = - {1 \over 2} \log (1 + \chi^{-}) \log (1 + {1 \over \chi^{+}})$.

The next piece is the critical value of Yang-Yang functional. In the case of $n=4$ the solution of the Y-system is simply $X_{1}(\zeta) = e^{Z_{1}/\zeta + \bar{Z}_{1} \zeta}$. Thus, we have
\eqn\YY{\eqalign{
YY_{cr} &= - {1 \over 2 \pi} \int_{l_{1}} d \theta {\cosh 2 \theta \over \sinh 2 \theta} ( \sinh \theta \log \chi^{+} - i \cosh \theta \log \chi^{-}) \log (1 + X_{1}(\theta)) \cr
&= \int_{- \infty}^{\infty} d t \frac{|m| \sinh t}{2 \pi \tanh (2 t + 2 i \phi)} \log (1 + e^{- 2 |m| \cosh t}), \quad \phi \in  (- {\pi \over 2}, 0)
}}
with
\eqn\Prob{\eqalign{
\log \chi^{+} &= -\log \chi^{+}_{2} =  |m| \cos \phi \cr
\log \chi^{-} &= -\log \chi^{-}_{2} = |m| \sin \phi. }}

The last step is $C_{0}(4) = {7 \pi \over 6}$ which finishes the calculation of the area for the octagon.
Notice that to get the answer in the form of \refs{\AldayYN} we should make change of variables $\phi = \tilde{\phi} - {\pi \over 2}$.

\appendix{F}{On the connection with BTZ black holes}

The cases which we have considered are very special cases of quotients of $AdS_3$ by
particular group element. All these are very special cases of BTZ black holes\foot{We are grateful to Benjamin Basso for pointing this out.} which are the solutions of $3$D Einstein gravity
\eqn\Ebh{\eqalign{
ds^2 &= - (r^2 - M) d \tau^2 - J dt d\phi+ \frac{d r^2}{r^2 - M + \frac{J}{4 r^2}} + r^2 d \phi^2 \cr
0 \leq &\phi \leq 2 \pi
}}
where we set $R=1$.

If we set $J=0$ and $M = - \frac{1}{m^2}$\foot{Although these solution have naked singularity we call them BTZ black holes in generalized sense \refs{\SteifZM}.} this geometry correspond to $AdS_3$ with the
identification $\phi \sim \phi + \frac{2 \pi}{m}$.
In the case of $m = \infty$ or $M=0$ this is the Poincare metric of $AdS_3$ with $\phi$ being spatial dimension and the
 identification
$\phi \sim \phi + 2 \pi$ which appears in the form factor problem. In the case $m=1$ this is the $AdS_3$ metric which appears in the amplitude problem.

One can consider the black holes with non-zero angular momenta and positive masses. However the physical interpretation of
these solutions in terms of field theory is not clear for us.

\appendix{G}{The monodromy in terms of coordinates}

Here we write formulas for $x^{+}$. Formulas for $x^{-}$ are completely the same.
As was explained in the text we can consider points $x^{+}_{i}$ with $i \in [0,n+2]$. Then we can use
conformal transformation to fix $x^{+}_{0},x^{+}_{1},x^{+}_{2}$. The monodromy relates coordinate of point $x^{+}_{i+n}$
to $x^{+}_{i}$. Thus, we can express the components of the monodromy in terms of coordinates $(x^{+}_{0},x^{+}_{1},x^{+}_{2},x^{+}_{n},x^{+}_{n+1},x^{+}_{n+2})$.
To do it we solve the following system of equations
\eqn\MonSol{\eqalign{
x^{+}_{i+n} &= {\hat{\Omega}_{11} + \hat{\Omega}_{12} x^{+}_{i} \over \hat{\Omega}_{21} + \hat{\Omega}_{22} x^{+}_{i}}, \quad i = 0,1,2 \cr
{\rm det} [\hat{\Omega}] &= 1
}}
where as usually when it is possible we enumerate the cusps in such a way that $x^{+}_{i+1} > x^{+}_{i}$.
The solution is
\eqn\MonSolution{\eqalign{
A &= x^{+}_{1,0} \ x^{+}_{2,0} \ x^{+}_{2,1} \ x^{+}_{n+1,n} \ x^{+}_{n+2,n} \ x^{+}_{n+2,n+1} \cr
\hat{\Omega}_{11} &= {1 \over \sqrt{A}} ( x^{+}_{0} x^{+}_{1} x^{+}_{n+1,n} + x^{+}_{1} x^{+}_{2} x^{+}_{n+2,n+1} - x^{+}_{0} x^{+}_{2} x^{+}_{n+2,n}) \cr
\hat{\Omega}_{22} &= {1 \over \sqrt{A}} ( x^{+}_{n} x^{+}_{n+1} x^{+}_{1,0} + x^{+}_{n+1} x^{+}_{n+2} x^{+}_{2,1} - x^{+}_{n} x^{+}_{n+2} x^{+}_{2,0}) \cr
\hat{\Omega}_{12} &= {1 \over \sqrt{A}} ( x^{+}_{2} x^{+}_{n+1,n} + x^{+}_{0} x^{+}_{n+2,n+1} - x^{+}_{1} x^{+}_{n+2,n}) \cr
\hat{\Omega}_{21} &= {1 \over \sqrt{A}} ( - x^{+}_{0} x^{+}_{n+1} x^{+}_{n+2} x^{+}_{2,1} +  x^{+}_{1} x^{+}_{n} x^{+}_{n+2} x^{+}_{2,0} - x^{+}_{2} x^{+}_{n} x^{+}_{n+1} x^{+}_{1,0}).
}}
As the monodromy is defined up to a sign we choose the solution by claiming that the case $x^{+}_{i+n} = x^{+}_{i}$ corresponds to the unit monodromy.

The trace of the monodromy is equal to
\eqn\Trace{
{\rm Tr}[\hat{\Omega}] = {1 \over \sqrt{A}} (x^{+}_{n+2,0} x^{+}_{n+1,n} x^{+}_{2,1} - x^{+}_{1,0} x^{+}_{n,2} x^{+}_{n+2,n+1}).
}
Notice that while the separate components of the monodromy are not conformal invariant the trace is as it should be.
The monodromy $\hat{\Omega}$ in this section corresponds to $\hat{\Omega}(\zeta = 1)$ in the main body of the text. For the $x^{-}$ coordinates it would be $\hat{\Omega}(\zeta = i)$.

\listrefs

\bye